\documentclass[12pt,preprint]{aastex}
\usepackage{amsmath}
\usepackage{amssymb}
\usepackage{natbib}

\shorttitle{F, G, K, M Spectral Standards in the $Y$ Band}
\shortauthors{Sharon et al.}
\begin{document}
\title{F, G, K, M Spectral Standards in the $Y$ Band (0.95-1.11 $\mu$m)}

\author{Chelsea Sharon \altaffilmark{1,2}}
\email{csharon@physics.rutgers.edu}
\author{Lynne Hillenbrand \altaffilmark{1}}
\email{lah@astro.caltech.edu}
\author{William Fischer \altaffilmark{3,4}}
\email{wfische@utnet.utoledo.edu}
\author{Suzan Edwards \altaffilmark{5}}
\email{sedwards@ast.smith.edu}
\altaffiltext{1}{Department of Astronomy, California Institute of Technology, Pasadena CA 91125, USA}
\altaffiltext{2}{current affiliation: Department of Physics and Astronomy, Rutgers The State University of NJ, Piscataway NJ 08854, USA}
\altaffiltext{3}{Department of Astronomy, University of Massachusetts, Amherst MA 01003, USA}
\altaffiltext{4}{current affiliation: Department of Physics and Astronomy, University of Toledo, Toledo, OH, 43606, USA} 
\altaffiltext{5}{Five College Astronomy Deptartment, Smith College, Northampton MA 01063, USA}

\begin{abstract} 
We take advantage of good atmospheric transparency and the availability of high quality instrumentation in the $1\, \micron$ near-infrared atmospheric window to present a grid of F, G, K, and M spectral standards observed at high spectral resolution ($R\approx$25,000).  In addition to a spectral atlas, we present a catalog of atomic line absorption features in the $0.95-1.11\,\micron$ range. The catalog includes a wide range of line excitation potentials, from 0-13 eV, arising from neutral and singly ionized species, most frequently those of Fe {\scshape i} and Ti {\scshape i} at low excitation, Cr\,{\scshape i}, Fe {\scshape i}, and Si {\scshape i} at moderate excitation, and C {\scshape i}, S {\scshape i}, and Si {\scshape i} having relatively high excitation.  The spectra also include several prominent molecular bands from CN and FeH. For the atomic species, we analyze trends in the excitation potential, line depth, and equivalent width across the grid of spectroscopic standards to identify temperature and surface gravity diagnostics near $1\, \micron$. We identify the line ratios that appear especially useful for spectral typing as those involving Ti\,{\scshape i} and C {\scshape i} or S {\scshape i}, which are temperature sensitive in opposite directions, and Sr\,{\scshape ii}, which is gravity sensitive at all spectral types. ASCII versions of all spectra are available in the online version of the journal.
\end{abstract}


\section{Introduction}\label{sec:intro}

Since the advent of infrared arrays in the early 1990s, there have been many low resolution spectroscopic atlases published covering the $J$, $H$, or $K$ bands.  Previous efforts that included the full O--M spectral classes and I--V luminosity classes made use of moderate resolution Fourier transform spectroscopy techniques \citep[e.\/g.\/,][]{Wall00,Mey98,Wall97}. An additional window of atmospheric transparency between these traditional near-infrared bands and the optical window, dubbed the $Y$ band, was highlighted by \citet{Hill02}. With existing observational capabilities on instruments at the IRTF, Keck, Las Campanas, and UKIRT telescopes, and with upcoming wide-field survey facilities such as Pan-STARRS and VISTA, the $Y$ band is becoming an increasingly appreciated and scientifically important near-infrared bandpass.  Several low resolution ($R\equiv\lambda/\Delta\lambda\sim2000$) spectral atlases for M, L, and T dwarfs have been published including $Y$ band data, e.g. \citet{Leg96}, \citet{McLe03}, and \citet{Cush05}.  However, a high-resolution spectroscopic atlas covering earlier spectral types and the full range of luminosity classes has not yet appeared. 

To complement existing moderate resolution catalogs in the $J$, $H$, and $K$ bands, we present here a high-resolution infrared spectral atlas for the $Y$ band.  Our data cover the $0.94-1.12\,\mu$m region sampled by NIRSPEC \citep{McLe98} at the W. M. Keck Observatory. We obtained spectra of 20 MK-classified stars ranging in spectral type from F through M and in luminosity class from I--V for use as spectroscopic standards in future work.  The data were processed as briefly described in Section \ref{sec:data}.  We present a representative grid of spectroscopic standards, a table of identified absorption features, and the resulting atlas of spectral lines in Section \ref{sec:atlas}.  In Section \ref{sec:analysis}, we present relevant atomic data and illustrate as well as discuss line sensitivity to stellar effective temperature and surface gravity.

\section{Target Selection, Data Acquisition, and Reduction}\label{sec:data}

Near-infrared echelle spectra of 20 stars with well-established MK classifications were acquired between 2001 and 2009 with NIRSPEC \citep{McLe98,McLe00} on Keck II.  The targets, listed in Table \ref{tab:stds}, were drawn largely from the lists of standards utilized in the \citet{Mey98} and \citet{Wall00} studies of stellar photospheres in the $H$ and $J$ bands, while the remainder were selected by SIMBAD \footnote{http://simbad.u-strasbg.fr/simbad/sim-fid} searches on spectral type and luminosity class, brightness, and accessibility. We used the N1 filter and have wavelength coverage from $0^{\prime\prime}.94$ to $1^{\prime\prime}.12$ $\mu$m at R = 25,000 ($\Delta$V = 12 km s$^{-1}$). The detector is a 1024 $\times$ 1024 pixel InSb chip, and the projected spectrograph slit we used was $0^{\prime\prime}.43\,\times\,12^{\prime\prime}$ in angular size.  The $Y$ band is covered in 12 full and 2 partial orders, without gaps.  The wavelength coverage of each order is given in Table \ref{tab:coverage}. 

Spectra were generally taken in nodded pairs using an ABBA pattern, with the astronomical target offset by $5^{\prime\prime}$ in each AB pair. Exposure times ranged from 1 to\,720 s, integrated in the case of each spectral standard to acquire signal-to-noise ratios $\approx100$ in the reduced one-dimensional spectra. Data reduction, including wavelength calibration and spatial rectification, extraction of one-dimensional spectra from the images, and removal of telluric emission and absorption features, is discussed in \citet{Edw06} and \citet{Fisch08}. We used a modified version of an IRAF script ``reduceit"
provided by Marianne Takamiya
\footnote{This was in conjunction with a 2001 Keck/NIRSPEC data set acquired as part of a Gemini--Keck time exchange.}
to reduce the 2002 and 2005 data, and we used the IDL package REDSPEC by S. Kim, L. Prato, and I. McLean to reduce data acquired in 2006 and 2009.

\section{The Spectral Atlas}\label{sec:atlas}
\subsection{Illustration}\label{sec:illustration}

From our 20 spectroscopic standards we identified a grid representing each of the four spectral classes (F, G, K, M)  and three luminosity classes (I-I-I, III, IV--V; hereafter supergiants, giants, dwarfs). The star used to typify each class is indicated in Table \ref{tab:stds} and was chosen based on having a good signal-to-noise ratio and relatively low rotational velocity ($v\,\sin(i)\lesssim20\,\text{km s}^{-1}$).
\footnote{HD\,21770 is the exception to this limit.  Of the two F III 
candidates it has the smaller rotational velocity.}
Figures \ref{fig:grid_80}--\ref{fig:grid_69} show the grid of representative spectroscopic standards displayed from shorter to longer wavelengths for orders 80 through 69. Each spectrum is normalized to the local continuum and shifted to line rest velocity. Orders 81 and 68 at the edges of our wavelength coverage are excluded due to the low signal-to-noise ratio and the relatively high telluric contamination towards the edges of the $Y$ band window. As can be seen, the 1 $\mu$m spectral region is rich in spectral features. The absorption lines originate from both atomic and molecular species.

\subsection{Line Identification}

In order to generate a complete list of the atomic and molecular lines seen in the grid of spectroscopic standards, we began with a candidate list of transitions from the Infrared Atlas of the Arcturus Spectrum \citep{Hink95}.  Arcturus is a K2 III star and so its associated line list should be relevant for the majority of the cool stars in our sample, with some supplementing needed as described below---especially for the hotter stars.  Of the 786 lines in our final candidate line list, 209 were added beyond those present in the Arcturus Atlas list. Because the Arcturus Atlas does not contain the complete details of each atomic transition (only line frequency and species), the candidate line list was cross-referenced with the Atomic Line Database \citep{Kurucz95} in order to obtain excitation potentials and log $gf$ values, as well as the Atomic Line List \citep{Peter}, to obtain the transition terms.  The Kurucz line list was developed from theoretical computations and it is far more extensive in line transitions (over 16,000 within our wavelength range) than necessary given the moderate spectral resolution and signal-to-noise ratio typical of our (and most) empirical astrophysical data. Thus great care was taken to properly match the atomic data to the lines identified in the Arcturus Atlas. Each line in the Arcturus Atlas was matched by species, ionization, and wavelength within $\pm~10^{-5}\,\micron$ to lines in the Kurucz Atomic Line Database (in vacuum wavelengths). In some cases this resulted in multiple potential matches between the lists among lines of the same species but having different ionization states.  Taking ionization into account generally resolved the situation but ambiguities often remained for lines such as the numerous Cr {\scshape i}, Fe {\scshape i}, and Ti {\scshape i} transitions. In such cases we chose to associate---at least initially---the Arcturus Atlas line with the Kurucz line possessing the largest log $gf$ among those with a realistic ($<$10\,eV) lower excitation potential. 

After the excitation potential, log $gf$, and transition term details of the atomic lines were added from the various sources to the candidate line list, the spectra were visually inspected.  We observed that some lines identified in the Arcturus Atlas did not correspond to any observed spectral features. Lines that were not present at $\lesssim 1\%$ of the continuum level were flagged appropriately (as having strength 0 on the scale of 0--3 described below), as were any multiply matched to those in the Kurucz list but decided not to be the correct identification.  
These lines are ignored in our further analysis but remain in the candidate line list for reference. 

Visual inspection of the spectra relative to the line list also identified strong absorption features seen in the spectra but not present in the assembled atomic line list. For later type stars (and especially prominent in the supergiants) such features were mostly identifiable as molecular lines based again on comparison to the Arcturus Atlas.  We do not list these individually herein, but we found that nearly all unidentified strong lines in the later type stars appear to be molecular.  For the earlier type F stars, additional atomic lines were identified using the Kurucz line list, now allowing for higher excitation potentials. In the case that there were multiple possible lines within the spectral resolution of the data ($0.4\,\text{\AA}$)
the line with lowest ground state energy level was chosen. If this line was a species that had not been previously identified in the spectra, then other line candidates near the observed line center were considered. 

After the initial matching using the Arcturus Atlas (Hinkle) and Atomic Line Database (Kurucz), all lines that were part of the same multiplet were identified using the on-line Atomic Line List (van Hoof). Any lines in multiplets of those already identified based on the above exercises and within the spectral range of the data were added to the line list if not already present, and are flagged as so in the final catalog.  As above, we considered only the expected range of log $gf$ and excitation potential values when initially assigning new candidate lines based on multiplets.  We then also considered the relative line strengths among the multiplets. Specifically, for optically thin (weak) lines in local thermodynamic equilibrium, the expected line intensity ratio $I_1/I_2$ is $A_1/A_2 ~ g_1/g_2 ~ \lambda_2/\lambda_1 ~ e^{\Delta E/kT}$  or $I_1/I_2 = g_1f_1/g_2f_2 ~ \lambda_2^3/\lambda_1^3 ~ e^{\Delta E/kT}$ where the symbols have their usual meanings.  For stronger lines the line broadening factors should be taken into account but we have ignored this aspect for ease of analysis.  In the case of multiplets (lines arising from the same lower level) the above expression simplifies because of the very small energy differences, to $I_1/I_2= (g_1f_1/g_2f_2)(\lambda_2/\lambda_1)^3$ which is what we use as a check on the line identifications. If during the multiplet analysis the case arose that an already-identified line could also be part of a multiplet from a different line species, we chose the multiplet identification over the alternative species unless both could be justified via multiplets or previous identification, in which case an additional flag for line blending was added. 

We note that the line wavelengths as well as their other data as given among the Arcturus Atlas (Hinkle), the CfA Atomic Line Database (Kurucz) and the on-line Atomic Line List (van Hoof) were not always identical within quoted errors.  Further, there are lines appearing in some of the above sources that are not available in the others. We believe sensible choices for line identifications have been made, and we have indicated remaining uncertainties in the tabulation.  We also note that in some cases we have been unable to discern the transition term from the existing sources of atomic data.  The 1 $\mu$m region is a relatively unscrutinized wavelength regime in stellar astrophysics and we are offering some of the first data that can help refine the available atomic line lists and parameters.

Finally, all lines in the catalog were assigned a numerical grade based on their relative strength in the spectroscopic standard grid. The grades range from zero to three, where the zero corresponds to lines that are not seen in any of the spectroscopic standards and are left in the catalog only for completeness, while the three is assigned to the deepest 2--10 lines per order (often deeper than 70\% of the continuum level across multiple spectral types and luminosity classes). Throughout this paper, we consider a one to be a weak line (shallower than 90\% of the continuum), while a two (deeper than 90\% of continuum) or a three (up to 50\% of continuum) is a strong line. Lines that were blended (at greater than the 50\% level), or lines that were only seen in the F standards were also noted in the catalog.  We note that for late-type and especially low-gravity stars, some expected atomic lines can be easily lost amidst strong molecular absorption features, generally CN lines which can exceed 80\% depth in some orders. The number of lines in each strength classification, and the fractions that are molecular, blended, or only seen in the F standards are indicated in Table \ref{tab:strengths}.

\begin{figure}\begin{center}
\includegraphics[width=6in, clip=true]{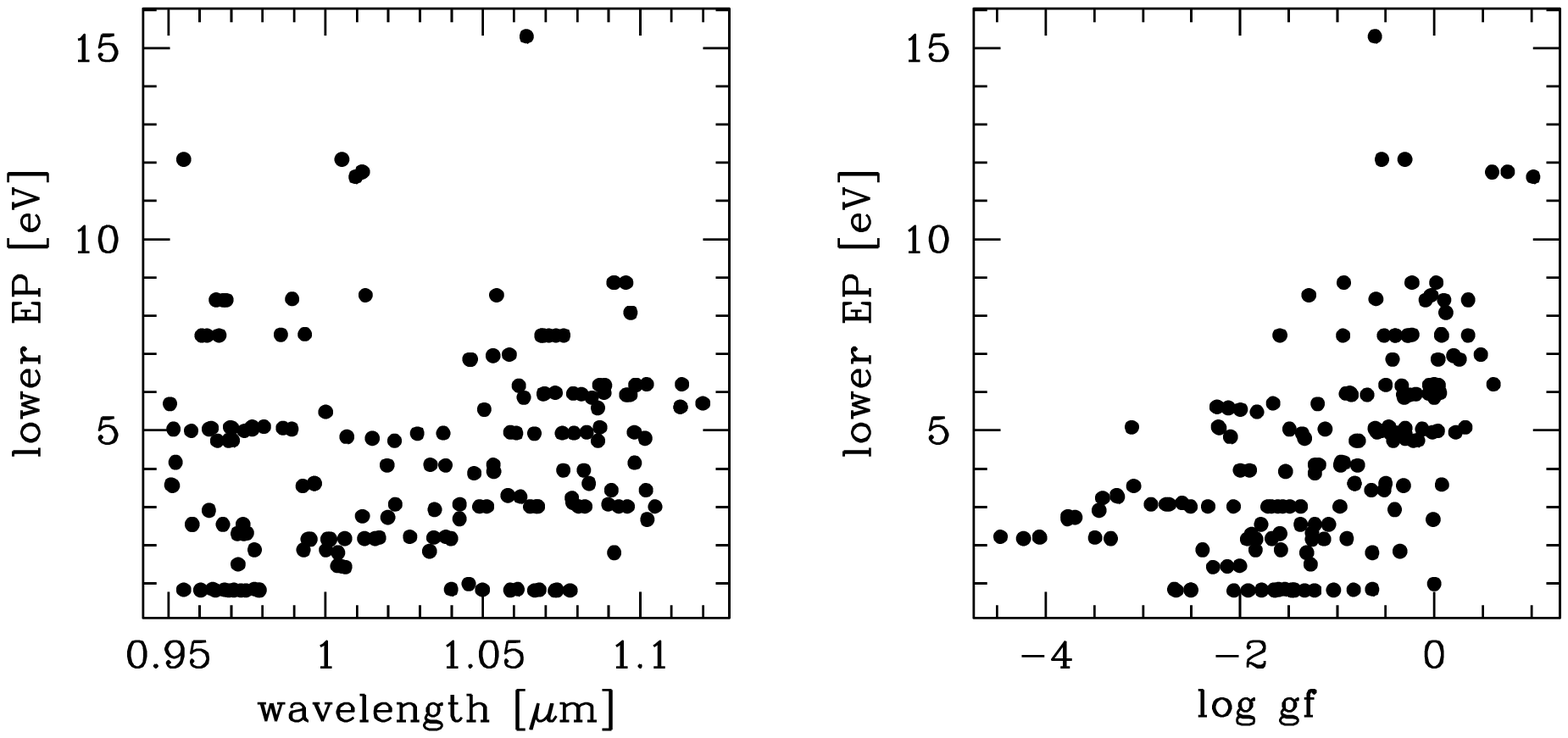}\end{center}
\figurenum{2}
\caption{Lower excitation potential as a function of wavelength (left) and
log $gf$ (right) for atomic lines identified in Table~\ref{tab:lines}. \label{fig:lowerEP}}
\end{figure}
\begin{figure}\begin{center}
\includegraphics[scale=0.65, clip=true]{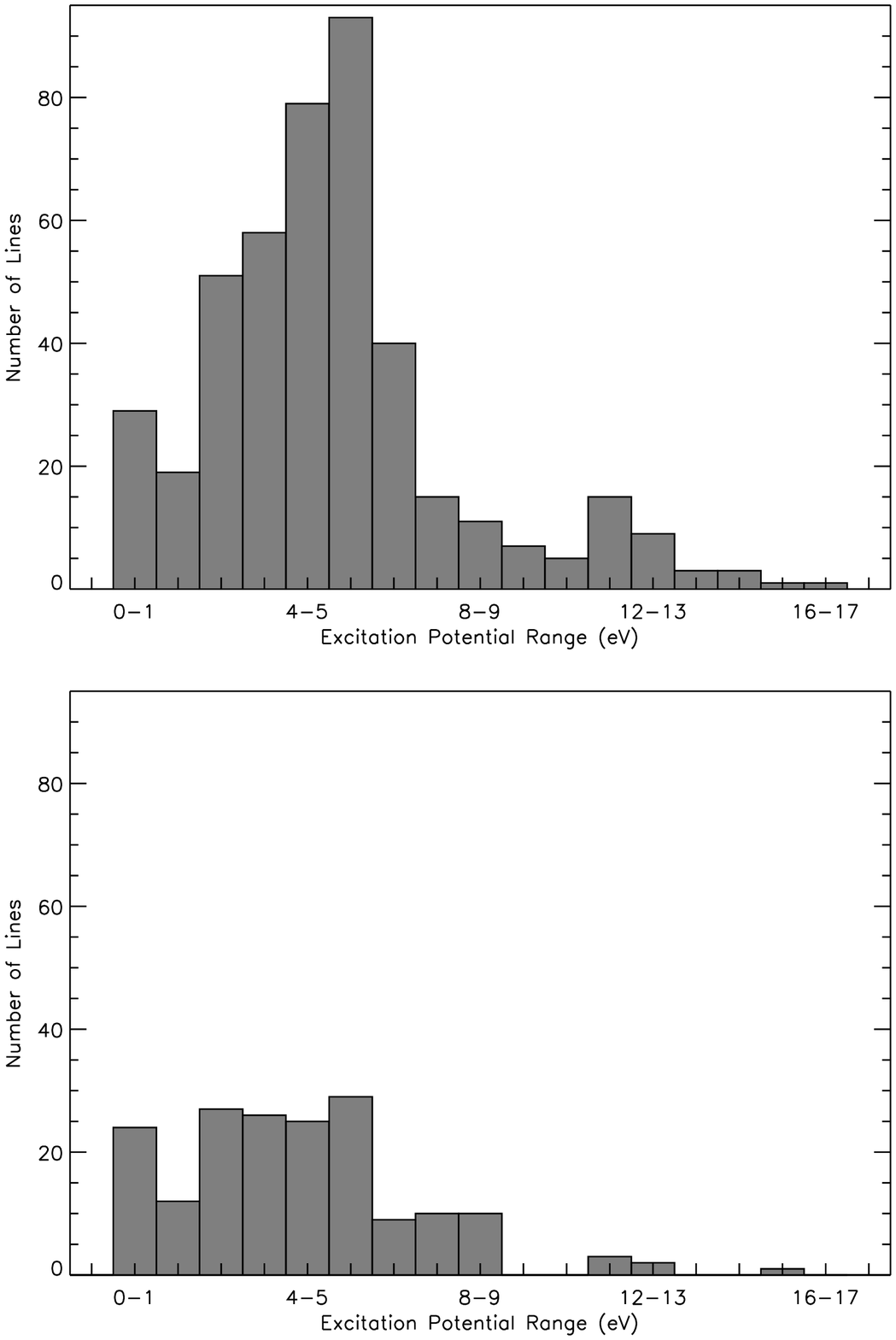}\end{center}
\figurenum{3}
\caption{Distribution of excitation potentials for all 453 identified atomic transitions (top panel) and those 188 categorized as strong lines with line depth exceeding $90\%$ of the continuum level and appearing in Table~\ref{tab:lines} (bottom panel). The lines are divided into $1\,{\rm eV}$ bins according to the energy of the lower transition level. Most lines are in the $0-8\,{\rm eV}$ range, but there is a tail extending to higher excitation potential, up to $14\,{\rm eV}$. While the strong lines are more uniformly distributed in the peak excitation potential range, they are still representative of the whole of the line list. 
\label{fig:ephisto}}
\end{figure}

The strongest (deepest) atomic and ionic lines, those given grades two and three above, plus their related atomic data are presented in Table \ref{tab:lines}.  The log $gf$ values among this strong line set range from $-4.5$ and up for Fe {\scshape i} and Ti{\scshape i} lines, to -2.5 and up for Cr {\scshape i} lines, to $-1.5$ and up for Si {\scshape i} lines, etc.  We show in Figure~\ref{fig:lowerEP} the lower excitation potential as a function of wavelength and of log $gf$ for these strong lines.  The general trends in  Figure~\ref{fig:lowerEP} are as expected from atomic physics considerations.  
In Figure~\ref{fig:ephisto}, we show a histogram of the number of atomic lines in 1\,eV bins of the lower excitation levels, for both the entire line catalog and for just the strong lines (those presented in Table \ref{tab:lines} and Figure~\ref{fig:lowerEP}).  This figure is discussed in more detail below. The table and figures include only the 0.95--1.11\,$\mu$m spectral region, where our final spectra are most uniform in signal-to-noise ratio across the standards grid.  

\section{Analysis}\label{sec:analysis}

The strong atomic lines identified above as well as the molecular lines are illustrated in Figures \ref{fig:fmgiant_pg1}--\ref{fig:fmdwarf_pg2} using the extremes of our spectral type grid (F and M, giants and dwarfs) for comparison.  Orders 79 and 71 appear to cover particularly useful spectral ranges for discriminating between hotter and cooler stars (see Section \ref{sec:diagnostic})
based on the trade off in strong features as a function of spectral type. 

Summarizing, notably present in the 1$\mu$m spectral region are numerous low and moderate excitation lines of neutral Ca {\scshape i}, Cr {\scshape i}, Fe {\scshape i}, K {\scshape i}, Mg {\scshape i}, Na {\scshape i}, Ni {\scshape i}, Si {\scshape i}, and Ti {\scshape i} along with singly ionized Sr {\scshape ii}. Lines of higher excitation potential like H {\scshape i} (and He {\scshape i}, though not within the effective temperature range sampled here), C {\scshape i}, N {\scshape i}, P {\scshape i}, and S {\scshape i} and singly ionized lines of Al {\scshape ii}, Ca {\scshape ii}, Fe {\scshape ii}, and Mg {\scshape ii} are also pervasive in the spectroscopic standards. Based on visual examination of the spectral grid, many of  these lines are temperature sensitive and a few, such as Sr {\scshape ii} ($1.003940,\,1.033010,\,1.091790\,\micron$), are gravity sensitive. Such trends among the atomic lines are discussed more quantitatively below.

The spectra also feature many molecular lines, including (for all G--K--M types) a very prominent CN 0--0 transition molecular series with band heads near 1.093 $\mu$m ($R_2$) and 1.097 $\mu$m ($R_1$) according to \citet{McKellar88}; these occur in the red part  of order 70 and the series covers much of order 69. At the later spectral types (M stars), the FeH molecule has a prominent 0--0 band head (the ``Wing--Ford" band which is apparent longward of $0.9896\,\micron$) and $Q$-branch (at $1.0061\,\micron$). 

At even later spectral types, for which we do not have data, the $Y$ band contains TiO bands which appear at 0.942 and 0.950 $\mu$m, FeH appearing at 1.025 and 1.064 $\mu$m \citep{Leg96}, and VO appearing at 0.971 and 1.064 $\mu$m through $1.08\,\micron$ \citep[]{Leg96,Cush03,Cush05}. There is also H$_2$O toward the short- and long-wavelength edges of the $Y$ band atmospheric window.  Late M and L dwarfs also feature very prominent atomic K {\scshape i}, Na {\scshape i}, and Ca {\scshape i}  features in the $Y$ band \citep{Leg96,2004A&A...416..655L} that are only weakly (grade one on our system) or not present (grade zero) yet in the latest types among our spectra.

\subsection{Trends in Excitation Potential}\label{sec:ex_pot}

We examine in Figure~\ref{fig:ephisto} the distribution of energy levels for the suite of atomic spectral features identified in the $Y$ band (complete candidate line list as well as the strong lines of Table~\ref{tab:lines}). The number of lines as a function of lower excitation potential is peaked at approximately 5 eV. The distribution declines more rapidly on the high excitation potential side of the peak than the lower, largely due to the presence of many strong moderate excitation Si {\scshape i} and Fe {\scshape i} transitions throughout the F, G, K, M spectral range. Heavier metals and higher ionization states generally contribute to the high excitation potential tail, while a large number of Ti {\scshape i} and some Cr {\scshape i} lines contribute to the lower excitation potential tail. Notable exceptions to this general trend include several C {\scshape i} lines at $\sim7-10$\,eV, and a few H {\scshape i} lines in the upper excitation potential range. As expected, many of these high excitation potential lines are only seen (with strength values of 3 or 2 on our system) among the earlier type F stars within our spectroscopic grid.

\begin{figure}\begin{center}
\includegraphics[scale=.55, clip=true]{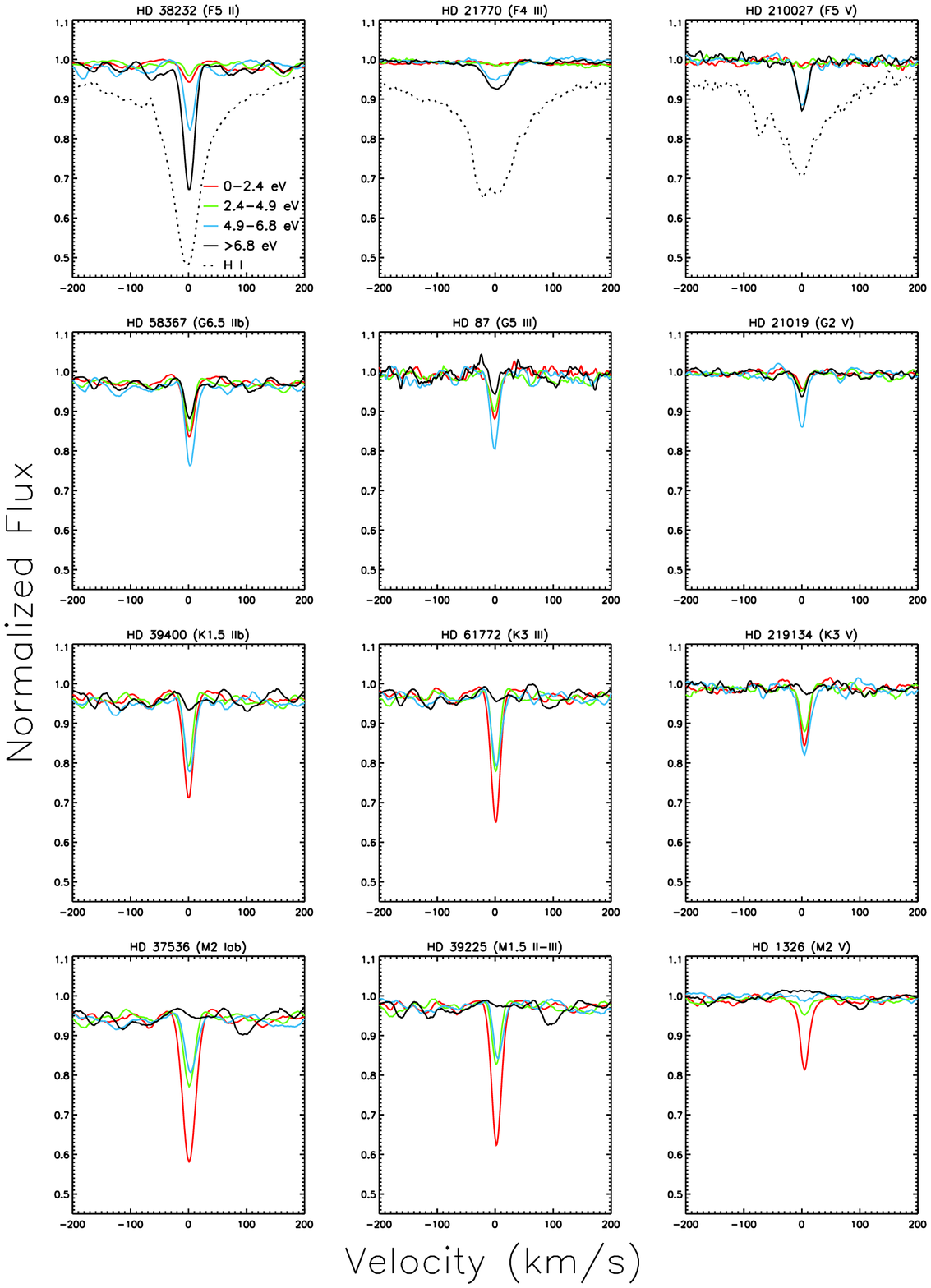}\end{center}
\figurenum{6}
\caption{Average profiles of strong lines in four different excitation potential ranges: 0--2.4\,eV (red; containing 35 lines), 2.4--4.9\,eV (green; containing 41 lines), 4.9--6.8\,eV (blue; containing 29 lines), and $>6.8\,{\rm eV}$ (black; containing 26 lines). The three H lines are excluded from the calculation of the solid black profile but are shown separately for the F stars as the dotted black line (in the cooler standards the H profile is too heavily blended with neighboring features for the absorption profile to be informative). Only lines that are unblended and strong enough to be tabulated in Table ~\ref{tab:lines} are included. A clear trend can be seen with hotter stars having stronger high-excitation lines compared to cooler stars having stronger low-excitation lines. Additionally, supergiants tend to have stronger lines then the dwarfs in the same spectral class. 
\label{fig:aveprofs}}
\end{figure}

In Figure~\ref{fig:aveprofs}, we have plotted for the spectroscopic standards grid the average line profile for four different bins in excitation potential. The atomic lines were divided into bins such that roughly the same number of $Y$ band lines are contained in each bin, but with an attempt to minimize the splitting of lines of the same species into multiple bins. This resulted in divisions at $2.4$, $4.9$, and $6.8\,{\rm eV}$. Only strong, unblended lines were included. However, hydrogen lines as well as any lines within $\pm 0.0005\, \micron$ of it were excluded since these particularly strong/broad lines dominate in the averaged profiles over nearby line features. There are 35 lines in the 0-2.4 eV bin, 41 lines in the 2.4--4.9 eV bin, 29 lines in the 4.9--6.8 eV bin, and 26 lines in the $>6.8$ eV bin. 

As expected, the hotter stars tend to have stronger lines with higher excitation potentials while cooler stars are dominated by low excitation potential lines. Supergiants also tend to have stronger lines overall relative to the dwarfs.
Specifically, the F stars are dominated by the 6.8\,eVand higher lines, the G stars are dominated by the 4.9--6.8\,eV lines, and the M stars are dominated by the 0--2.4\,eV lines. One would expect that the K stars would lie in sequence and be dominated by the 2.4--4.9\,eV lines. Instead, we find that the line strengths are roughly constant from 0 to 6.8\,eV and comparable to the 4.9--6.8\,eV bin in the G standards.  Although the K standards do fit within the overall excitation-potential-temperature trend, they do not match the observed pattern in detail in the expected way. 

\subsection{Diagnostic Lines for Spectral Classification}\label{sec:diagnostic}

Among the strong lines there are some obvious temperature dependencies apparent in Figures~\ref{fig:fmgiant_pg1}-\ref{fig:fmdwarf_pg2}.  For example, in order 79, there are several strong C {\scshape i} lines that appear in the early spectral types and are not seen in the later types. In the same order there are also Ti\,{\scshape i} lines that complement this trend, appearing very strongly in the late spectral types, but only weakly, if at all, in the early spectral types.  A similar trend can be seen in order 73 for Si {\scshape i} relative to Ti {\scshape i}, and in order 71 for both C {\scshape i} and Si {\scshape i} relative to Ti {\scshape i}.

\begin{figure}\begin{center}
\includegraphics[scale=.55, clip=true]{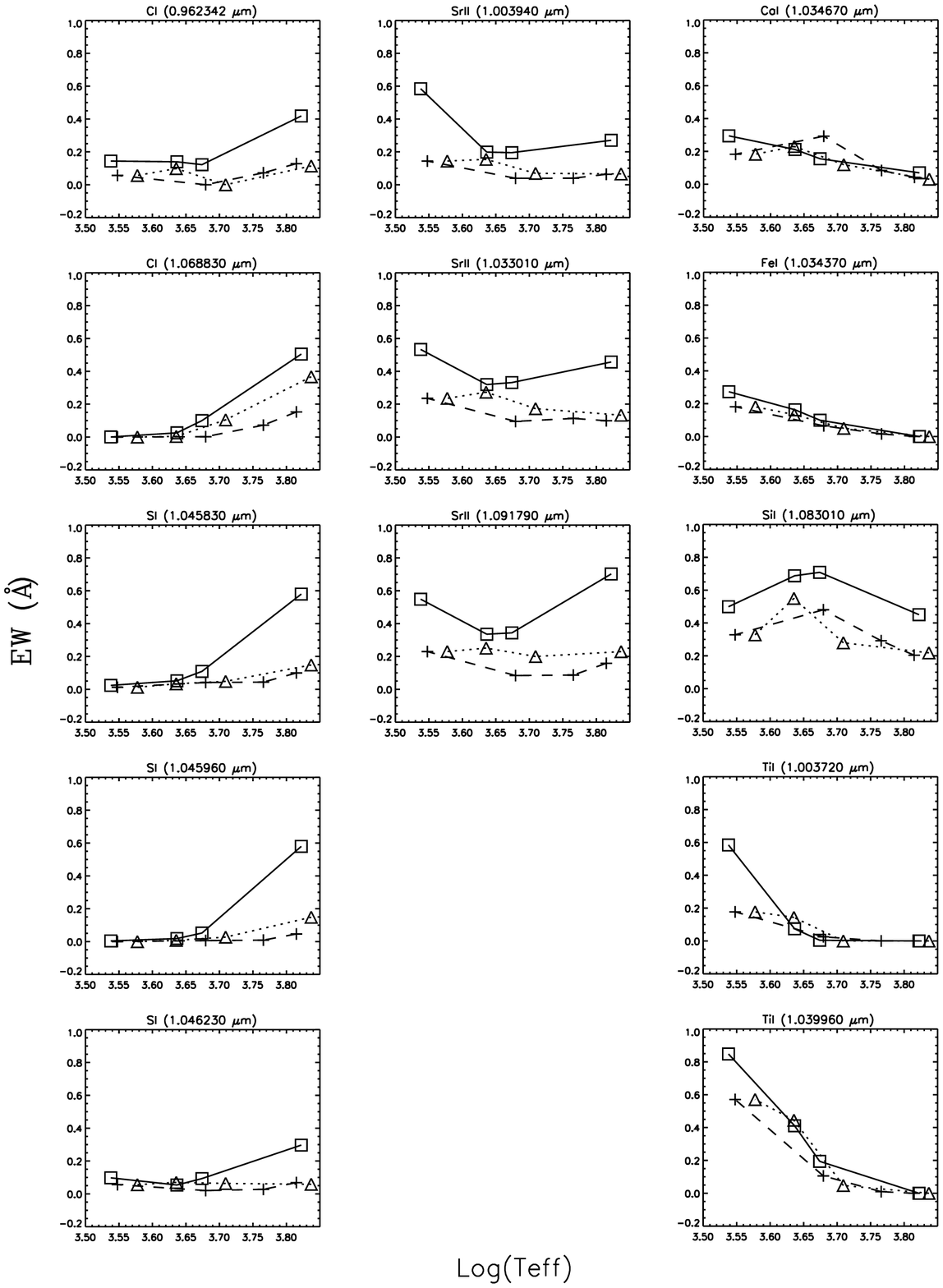}\end{center}
\figurenum{7}
\caption{Equivalent widths in angstroms of diagnostically useful lines vs. the effective temperature of the spectroscopic standards.  Squares represent supergiants, triangles the giants, and crosses the dwarf standards.   Lines connect the existing data and should not be misinterpreted as defining the detailed trend with temperature. Lines were chosen from Table~\ref{tab:lines} as those of strength three with significant temperature and/or gravity sensitivity within our spectral grid.  Generally, only one line from among those in a multiplet set is shown; its behavior is typical of all lines in the multiplet. The three columns depict lines with higher excitation potential (left), lines with lower excitation potential (right), and lines sensitive to surface gravity (center). Within each column the panels are ordered from top to bottom alphabetically by line species and then by increasing wavelength. 
\label{fig:equivwidths}}
\end{figure}

We investigated through equivalent width analysis the behaviors with both effective temperature and surface gravity of the various strong lines in the 1 $\mu$m region.  Several strong lines stand out over the others in Table~\ref{tab:lines} as potentially useful for analytic or diagnostic purposes, the strongest of which have line depths greater than 50\% of the continuum level. Figure~\ref{fig:equivwidths} plots the equivalent widths of several such lines versus the standard star effective temperature.  In constructing this plot we were particularly sensitive to including those lines with temperature dependencies discernable for dwarf stars.  As discussed above, lines of note include C {\scshape i} and S {\scshape i} multiplets, which show a strong temperature dependence, and several Ti\,{\scshape i} multiplets, which show a strong inverse temperature dependence.  This behavior was also highlighted by \citet{2004A&A...416..655L} for the a5F--z5Go multiplet of Ti {\scshape i} in the 1.04-1.07\,$\mu$m region, though we find many of these long wavelength lines of the multiplet blended with other strong lines.  Ca {\scshape i}, Cr {\scshape i}, and Fe {\scshape i} have generally weaker (inverse) temperature dependence. Several neutral H {\scshape i} lines and higher excitation potential neutral metals such as C {\scshape i}, P {\scshape i}, and S {\scshape i} and the singly ionized metals Ca {\scshape ii}, Fe {\scshape ii}, and Mg {\scshape ii} appear only in the F standards and are therefore also good temperature indicators.  The Si {\scshape i} lines peak in strength in the G--K spectral range and become weaker toward both F and M types, with some gravity sensitivity. In addition to their temperature dependence, the Ti {\scshape i} lines are also gravity sensitive at log $T_{\rm eff} <$ 3.7 as are the C {\scshape i} and S {\scshape i} lines at log $T_{\rm eff}\,>3.7$. Sr\,{\scshape ii} shows gravity dependence at all temperatures but little trend with temperature. 

\begin{figure}\begin{center}
\includegraphics[scale=.6, clip=true]{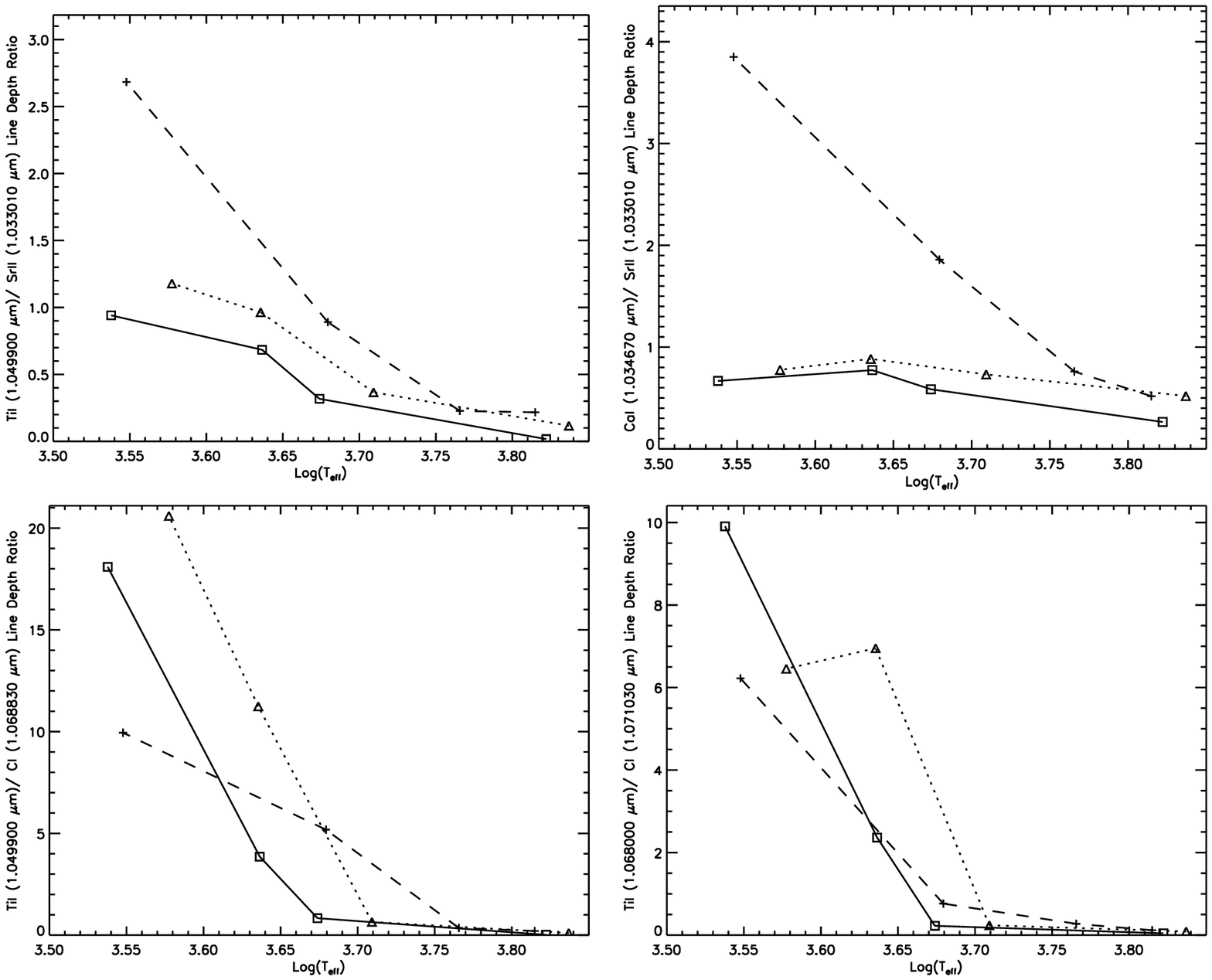}\end{center}
\figurenum{8}
\caption{Top panels: the ratio of line strengths for a temperature-sensitive and a gravity-sensitive line, as a function of the stellar effective temperature. The left side shows  Ti {\scshape i} ($1.049900\,\mu {\rm m}$) $/$ Sr {\scshape ii} ($1.033010\,\mu {\rm m}$), the lines with the strongest temperature and gravity trends in our spectra, while the right side shows  Ca {\scshape i} ($1.034670\,\mu {\rm m}$) $/$ Sr {\scshape ii} ($1.033010\,\mu {\rm m}$) which are chosen due to their proximity in wavelength. Bottom panel: the ratio of line strengths for two temperature sensitive lines with opposite temperature dependence, Ti {\scshape i} ($1.049900\,\mu {\rm m}$) and C {\scshape i} ($1.068830\,\mu {\rm m}$) on the left side and  Ti {\scshape i} ($1.068000\,\mu {\rm m}$) and C {\scshape i} ($1.071030\,\mu {\rm m}$) on the right side, as a function of stellar effective temperature. Squares represent supergiants, triangles the giants, and crosses the dwarf standards. Although strong dependences are evident, the illustrated lines connect the existing data only and should not be misinterpreted as defining the detailed trends with temperature.  
\label{fig:diagnostic}}
\end{figure}

We have identified a few lines having line depth ratio that are useful indicators of spectral and luminosity class. These ratios versus standard star effective temperature are illustrated in Figure~\ref{fig:diagnostic}. The Ti {\scshape i} ($1.049900\,\micron$) and Sr {\scshape ii} ($1.033010\,\micron$) lines, are the temperature and gravity sensitive lines, respectively, with the strongest trends apparent in  Figure~\ref{fig:equivwidths}. In this case, division of the temperature-sensitive line by the gravity-sensitive line spreads the spectroscopic standards such that dwarfs and then giants have increased line strength relative to supergiants at similar effective temperatures. The Ti {\scshape i} ($1.049900\,\micron$) and  C {\scshape i} ($1.068830\,\micron$) lines are both temperature sensitive, but the temperature dependence of C {\scshape i} is opposite that of Ti {\scshape i}. This creates an amplified trend in line strength versus temperature relative to that of the single lines 
(as in Figure~\ref{fig:equivwidths}). We also investigated lines pairs that were closer together in wavelength and therefore suitable for temperature and gravity designation from single-order moderate to high dispersion spectra.  Thus the Ca\,{\scshape i} ($1.034670\,\mu {\rm m}$) $/$ Sr {\scshape ii} ($1.033010\,\mu {\rm m}$) ratio is also illustrated as sensitive to gravity and the Ti {\scshape i} ($1.068000\,\mu {\rm m}$) $/$ C\,{\scshape i} ($1.071030\,\mu {\rm m}$) ratio as sensitive to temperature. Either of these line pair sets could be used in comparing the relative spectral typing of a set of previously unclassified stellar spectra, depending on the spectral dispersion

Although not present in our effective temperature range, there is a high excitation (21\,eV) He {\scshape i} line at $1.08333\,\micron$ in addition to the high excitation potential (12 eV) H {\scshape i} lines already noted here.  These He {\scshape i} and H {\scshape i} lines are useful spectral type diagnostics for O, B, and A stars.  Both lines are also particularly sensitive probes at all spectral types of energetic circumstellar processes such as chromospheric activity and wind processes \citep{Edw06,Fisch08}.

\section{Summary}\label{sec:summary}

We have presented a spectral atlas for the $Y$ band region of the near infrared portion of the electromagnetic spectrum.  Our data set consists of a grid of MK-classified stars for use as spectroscopic standards spanning the F through M spectral classes and I--V luminosity classes observed at R$\approx$25,000. The associated catalog of identified lines contains numerous spectral absorption features, both atomic and molecular in origin, that span a wide range of excitation potentials. Many of the atomic features show strong line depth and equivalent width trends across the spectral atlas. This atlas and the identified atomic spectral lines are potentially useful for two-dimensional spectral classification,
especially line depth ratios involving Ti {\scshape i} and C {\scshape i} or S {\scshape i}, which are temperature sensitive in opposite directions, and Sr {\scshape ii}, which is gravity sensitive. ASCII versions of all spectra are available to download with the online version of the journal (an example of the form and content of these spectra can be found in Table \ref{tab:asciiex}).

\acknowledgments

We thank the anonymous referee for his/her careful inspection of the manuscript.
We acknowledge with gratitude the builders of NIRSPEC and the hard work and assistance from the staff at W.M. Keck Observatory. The observatory is operated as a scientific partnership among the California Institute of Technology, the University of California and the National Aeronautics and Space Administration and was made possible by the generous financial support of the W.M. Keck Foundation.  
The authors wish to recognize and acknowledge the very significant cultural role and reverence that the summit of Mauna Kea has always had within the indigenous Hawaiian community.  We are most fortunate to have the opportunity to conduct observations from this mountain.  
Some of the spectra presented herein were obtained by Dan Stark and  Richard Ellis or Colette Salyk and Geoff Blake; we are grateful for their contributions.   
We consulted The Atomic Line List
(http://www.pa.uky.edu/$\sim$peter/atomic/) 
and the CfA Kurucz line list 
(http://cfa-www.harvard.edu/amp/ampdata/kurucz23/sekur.html)
while conducting this research.
We are grateful to Andrew Baker for his tolerance during the final analysis and writing stages of this paper.

\begin{deluxetable}{llrclrc}
\tabletypesize{\scriptsize}
\setlength{\tabcolsep}{0.03in}
\tablewidth{0pt}
\tablecaption{Spectral Reference Stars \label{tab:stds}}
\tablehead{
{(1)}        &{(2)}   & {(3)} & {(4)}   & {(5)}  & {(6)}  &{(7)}\\
{Name}  &{Alt. Name} &  {Obs. Date} & {Sp. Type}& $T_{eff}$ [K]\tablenotemark{a} & Sp. Type Reference & Template\tablenotemark{b}} 
\startdata
HD 21770 & HR 1069 &	    2006 Nov 30&     F4 III &	7170&	\citet{Rom52}	 & X \\
HD 193370 & HR 7770 &	    2006 Nov 30&     F5 Ib &	6640&	\citet{Griff60}	&  \\
HD 38232 & BD+29 983 &   2005 Dec 13&     F5 II &		6640&	\citet{Bidel57}	& X \\
HD 55052 & HR 2706 &	    2002 Nov 26&     F5 III &	6470&	\citet{Harlan69}	 &  \\
HD 210027 & HR 8430 &	    2006 Nov 30&     F5 V &	6530& 	\citet{MKatlas43}	& X \\
HD21019	& HR 1024 &        2006 Nov 29&    G2 V &          5830&       \citet{Harlan74} & X  \\
HD 87    & HR 4 &	             2009 Jul 11&     G5 III & 	        5072\tablenotemark{c}   &	\citet{Barnes86} & X \\
HD 58367 & HR 2828 &	    2005 Dec 13&     G6 IIb &	4720\tablenotemark{c}&	\citet{Keen89} & X \\
HD 75935 & BD+27 1682 & 2006 Nov 30&     G8 V&	5430\tablenotemark{c}&	\citet{Montes01}  &  \\
HD 166 & HR 8    &		     2006 Nov 30&     K0 V&	5240&	\citet{Basri90}  &  \\
HD 39400 & HR 2037 &	    2005 Dec 13&     K2 IIb&	4330&	\citet{Keen89}	& X \\
BD +59 2667& \nodata &	2006 Nov 30&     K2 IV&		5010&	\citet{Basri90}	&  \\
HD 61772 & HR 2959 &	2005 Dec 13&     K3 III &		4320&	\citet{Houk88}	 & X \\
HD 219134 & HR 8832 &	2006 Nov 30&     K3 V&		4780\tablenotemark{c}&  \citet{MKatlas43} & X \\
HD 201091 & 61 CygA & 2006 Dec 01&     K5 V&		4340&	\citet{MKatlas43}	&  \\
HD 201092 & 61 CygB & 2006 Dec 01&     K7 V&		4040& 	\citet{Griff60}	&  \\
HD 232979 & BD+52 857 & 2006 Nov 30&     M0.5 V&	5240&	\citet{Keen89}	&  \\
HD 39225& HR 2028 &	2005 Dec 13&     M1+ III &	      3550&         \citet{Keen89}	& X \\
HD 37536& HR 1939 &	2005 Dec 13&     M2 Iab&	       3450&	\citet{Sharp66}	& X \\
HD 1326 & Gl 15A  &   	2006 Nov 30&     M2 V&		3530&	\citet{Keen89}	 & X \\
\enddata
\tablenotetext{a}{Effective temperature as given by \citet{Mey98}.}
\tablenotetext{b}{Star was chosen to represent its temperature and luminosity class in the grid of spectroscopic standards.}
\tablenotetext{c}{No effective temperature listed in \citet{Mey98}. This value is an interpolation.}
\end{deluxetable}

\pagebreak
\begin{deluxetable}{lrcl}
\setlength{\tabcolsep}{0.03in}
\tablewidth{0pt}
\tablecaption{Effective Wavelength Coverage \label{tab:coverage}}
\tablehead{
{(1)}        &{(2)} \\
{Order}  &{$\lambda$ Range [$\mu$m]}} 
\startdata
81	&	0.94118 - 0.94813	\\
80	&	0.94646 - 0.95985	\\
79	&	0.95835 - 0.97192	\\
78	&	0.97054 - 0.98543	\\
77	&	0.98304 - 0.99696	\\
76	&	0.99590 - 1.00998	\\
75	&	1.00906 - 1.02332	\\
74	&	1.02255 - 1.03709	\\
73	&	1.03628 - 1.05122	\\
72	&	1.05081 - 1.06578	\\
71	&	1.06553 - 1.08065	\\
70	&	1.08065 - 1.09604	\\
69	&	1.09584 - 1.11184	\\
68	&	1.11241 - 1.12224	\\
\enddata
\end{deluxetable}

\pagebreak
\begin{deluxetable}{cccccccc}
\tabletypesize{\scriptsize}
\tablewidth{0pt}
\tablecaption{Distribution of Line Strengths \label{tab:strengths}}
\tablehead{
{(1)}        &{(2)}  & \multicolumn{2}{c}{(3)} & \multicolumn{2}{c}{(4)} & \multicolumn{2}{c}{(5)}\\
{Strength Class} & {Line Depth\tablenotemark{a}} &  \multicolumn{2}{c}{No. Lines\tablenotemark{b}}  & \multicolumn{2}{c}{No. in F Stars Only\tablenotemark{c}} & \multicolumn{2}{c}{No. in Blends\tablenotemark{d}} \\   
{} & {} & {Atomic} & {Molecular} & {Atomic} & {Molecular} & {Atomic} & {Molecular} }
\startdata
0 & Unobserved&  99 &\nodata & \nodata & \nodata & \nodata & \nodata \\
1 & $>90\%$ & 237 &279&  20 &0  & 2 &0 \\
2 & $<90\%$, $\gtrsim70\%$ & 141 &50 & 18 &0 & 11 &22 \\
3 & $\lesssim70\%$ & 75 &4 & 14 &0 & 10 &0 \\
\enddata
\tablenotetext{a}{As measured by minimum percentage of continuum level.}
\tablenotetext{b}{Number of atomic and molecular lines in this strength class.
The strength zero lines are either those from the  Arcturus Atlas which we can not identify in our lower resolution spectra, or those added to the candidate line catalog based on the multiplet analysis but not present in our data}.
\tablenotetext{c}{Number of atomic and molecular lines in this strength class that are only seen in the F class spectroscopic standards.}
\tablenotetext{d}{Number of atomic and molecular lines in this strength class that are blended with others at least the $50\%$ level.}
\end{deluxetable}

\begin{deluxetable}{lcrrrllc}
\tabletypesize{\scriptsize}
\setlength{\tabcolsep}{0.03in}
\tablewidth{0pt}
\tablecaption{Identified Atomic Spectral Lines \label{tab:lines}}
\tablehead{
{(1)}            &{(2)}   & {(3)}       & {(4)}           & {(5)}          & {(6)}          & {(7)}  & {(8)}\\
{Wavelength[$\micron$]} & {Atom/Ion} &   {log $gf$ }   & {$E_{lower}$ [eV]}  & {$E_{upper}$ [eV]} & {Term} & {Strength} & {Comment} } 
\startdata
0.9504665&       Mn {\scshape i}&  -1.200&   5.69633&   7.00088& ??      &       2B & Uncertain but in the Arcturus list  \\
0.950867&       Ti {\scshape i}&   0.078&   3.58290&   4.88689&y5Fo-f5F &     2B   &\\
0.951329&       Ti {\scshape i}&  -0.316&   3.55869&   4.86196&y5Fo-f5F &     2B  & \\
0.951587&       Fe {\scshape i}&  -1.491&   5.03341&   6.33641& x5Fo-f5F&          2 &  \\
0.952265&       N {\scshape ii}&  -0.960&   4.16750&   5.46958&y3Do-e3D &          2  & \\
0.954859&        H {\scshape i}&  -0.540&  12.08751&  13.38597& 3-8     &     3F  & \\
0.954865&       Ti {\scshape i}&  -1.598&   0.83605&   2.13459&a5F-z5Fo &          3 & * \\
0.957253&       Fe {\scshape i}&  -0.542&   4.99161&   6.28690&x5Fo-e5G &          2 & \\
0.957438&       Cr {\scshape i}&  -1.784&   2.54444&   3.83949&a5G-z5Fo &         2B & \\
0.957692&       Cr {\scshape i}&  -1.088&   2.54479&   3.83949&a5G-z5Fo &     3B & \\
0.960218&       Ti {\scshape i}&  -1.440&   0.82591&   2.11721&a5F-z5Fo &     3 & * \\
0.960567&        C {\scshape i}&  -0.940&   7.48090&   8.77172&3Po-3S   &     3F & \\
0.962342&        C {\scshape i}&  -0.516&   7.48278&   8.77114&3Po-3S   &     3F & \\
0.962903&       Cr {\scshape i}&  -3.451&   2.91531&   4.20301&a3P-z3Po &     2B & \\
0.962913&       Fe {\scshape i}&  -0.610&   5.03340&   6.32108&x5Fo-e5G &        2 & \\
0.963682&       Fe {\scshape i}&  -0.607&   5.06412&   6.35078&x5Fo-e5G &        2 & \\
0.964090&       Ti {\scshape i}&  -0.638&   0.84848&   2.13459&a5F-z5Fo &         3B &Only weak blend, * \\
0.965002&       Ti {\scshape i}&  -1.462&   0.81820&   2.10309&a5F-z5Fo &        3& *  \\
0.965222&        S {\scshape i}&   0.350&   8.41201&   9.69661&3Do-3D   &     2F & \\
0.965576&       Fe {\scshape i}&  -0.778&   4.73346&   6.01759&y3Do-e3F &         2 & \\
0.966108&        C {\scshape i}&  -0.264&   7.48830&   8.77172& 3Po-3S  &         3F & \\
0.967314&       Cr {\scshape i}&  -1.230&   2.54400&   3.82626&a5G-z5Fo &          2 & \\
0.967494&        S {\scshape i}&  -0.090&   8.40872&   9.69031&3Do-3D   &         2F & \\
0.967815&       Ti {\scshape i}&  -0.831&   0.83605&   2.11721&a5F-z5Fo &          3& *  \\
0.968346&        S {\scshape i}&   0.100&   8.40982&   9.69028&3Do-3D   &     2F & \\
0.969133&       Ca {\scshape i}&  -0.422&   4.73889&   6.01830&3Do-3P   &         2B &  \\
0.969146&       Ti {\scshape i}&  -1.638&   0.81300&   2.09240&a5F-z5Fo &          3& *  \\
0.969651&      Fe {\scshape i}]&  -3.115&   5.08593&   6.36467&x5Fo-e7F &         2B&  \\
0.970242&      Fe {\scshape i}]&  -2.215&   5.06412&   6.34208&x5Fo-e7F &         2 &\\ 
0.970459&       Ca {\scshape i}&  -0.169&   4.74385&   6.02151&3Do-3P   &          2 & \\
0.970834&       Ti {\scshape i}&  -1.037&   0.82591&   2.10309&a5F-z5Fo &          3& *  \\
0.971964&       Ti {\scshape i}&  -1.590&   2.30552&   3.58121&z5Do-a5D &          2&  \\
0.972167&       Ti {\scshape i}&  -1.272&   1.50259&   2.77801&a1G-z1Fo &          2&  \\
0.973100&       Ti {\scshape i}&  -1.234&   0.81820&   2.09240&a5F-z5Fo &          3& *  \\
0.973719&       Cr {\scshape i}&  -1.377&   2.54400&   3.81740&a5G-z5Fo &          2 & \\
0.974037&       Ti {\scshape i}&  -1.882&   2.29697&   3.56995&z5Do-a5D &     2B & \\
0.974124&       Fe {\scshape i}&   0.037&   4.99161&   6.26447&x5Fo-e5G &         2B & \\
0.974625&       Ti {\scshape i}&  -1.334&   0.81300&   2.08521&a5F-z5Fo &          3 &*  \\
0.974956&       Ti {\scshape i}&  -1.258&   2.31807&   3.58984&z5Do-a5D &         2B & \\
0.976610&       Fe {\scshape i}&  -0.468&   5.09989&   6.36951&x5Fo-e5G &         2B & \\
0.976658&       Fe {\scshape i}]&  -1.123&   5.03341&   6.30297&x5Fo-e7F &         2B&  \\
0.977293&       Ti {\scshape i}&  -1.604&   0.84848&   2.11721&a5F-z5Fo &          3&   \\
0.978602&       Ti {\scshape i}&  -1.452&   0.83605&   2.10309&a5F-z5Fo &          3& * As blend  \\
0.978623&       Ti {\scshape i}&  -1.643&   0.81820&   2.08521&a5F-z5Fo &     3& * As blend  \\
0.979030&       Ti {\scshape i}&  -1.469&   0.82591&   2.09240&a5F-z5Fo &          3& * Blend  \\
0.980352&      Fe {\scshape i}]&  -2.223&   5.09989&   6.36467&x5Fo-e7F &          2&\\
0.985746&     Ca {\scshape ii}&  -0.228&    7.505644&    8.763499&2Po-2S & 2F&\\
0.986444&       Fe {\scshape i}&  -0.299&   5.06412&   6.32109&x5Fo-e5G &         2F& \\
0.989175&       Fe {\scshape i}&  -0.125&   5.03341&   6.28690&x5Fo-e5G &         2 & \\
0.989334&      Ca {\scshape ii}&  -0.601&    8.438551&    9.69184&2Fo-2G   &  3F&\\
0.992711&      Fe {\scshape i}]&  -3.095&    3.54676 &    4.79579&a1D-y3Do   & 2&\\
0.993006&       Ti {\scshape i}&  -1.580&   1.87933&   3.12799&a3G-y3Fo &          2 &\\
0.993410&     Ca {\scshape ii}&  0.072 &   7.515348 &   8.763499 &2Po-2S & 2F&\\
0.994412&      Ti {\scshape i}&   -1.924&   2.16043&   3.40732&a3D-x3Do &  2  &   \\
0.995181&       Ti {\scshape i}&  -1.831&   2.15370&   3.39963&a3D-x3Do &         2B& \\
0.996401&       Na {\scshape i}&  -0.820&   3.61721&   4.86162&2D-2Fo   &         2B& Doublet, $\dag$\\
1.000032&      Fe {\scshape ii}&  -1.826&   5.48450&   6.72439&z4Fo-b4G &     2F &\\
1.000070&       Ti {\scshape i}&  -1.840&   1.87329&   3.11313&a3G-y3Fo &         2B& \\
1.000580&       Ti {\scshape i}&  -1.135&   2.16043&   3.39963&a3D-x3Do &          2& \\
1.001460&       Ti {\scshape i}&  -1.258&   2.15370&   3.39182&a3D-x3Do &          2& \\
1.003720&       Ti {\scshape i}&  -2.002&   1.46024&   2.69556&b3F-z3Go &          3& \\
1.003940&      Sr {\scshape ii}&  -1.312&   1.80482&   3.03988& 2D-2Po  &          3& \\
1.005160&       Ti {\scshape i}&  -2.131&   1.44325&   2.67681&b3F-z3Go &        3 &\\
1.005210&        H {\scshape i}&  -0.303&  12.08830&  13.32180&  3-7    &         3F &\\
1.006050&       Ti {\scshape i}&  -0.900&   2.17485&   3.40732&a3D-x3Do &          2 &\\
1.006270&       Ti {\scshape i}&  -2.278&   1.42985&   2.66206&b3F-z3Go &          2 &\\
1.006780&      Fe {\scshape i}&   -2.098&   4.83518&   6.06676&y3Do-e3F &  2   & \\
1.009498&      Mg {\scshape ii}&   1.020&  11.63047&  12.85873&2Fo-2G   &     2F &\\
1.0115253&       N {\scshape i}&   0.595&  11.75833&  12.98412& 4Do-4F &   2F&\\
1.011680&       Fe {\scshape i}&  -3.771&   2.75876&   3.98438&a3G-z3Fo &          2& \\
1.0117413&       N {\scshape i}&   0.755&  11.76464&  12.990176&4Do-4F  & 2F&\\
1.012370&       Ti {\scshape i}&  -1.672&   2.17485&   3.39963&a3D-x3Do &         2B& \\
1.012660&        C {\scshape i}&  -0.032&   8.53767&   9.76209& 1P-1Po  &         3F& \\
1.014830&       Fe {\scshape i}&  -0.298&   4.79579&   6.01759&y3Do-e3F &          2& \\
1.015790&       Fe {\scshape i}&  -4.229&   2.17609&   3.39674&a5P-z5Fo &          2& \\
1.017030&       Fe {\scshape i}&  -4.061&   2.19801&   3.41718&a5P-z5Fo &          2& \\
1.017326&       Ti {\scshape i}&  -3.292&   1.44325&   2.66206&b3F-z3Go&           2& \\
1.019192&       Ti {\scshape i}&  -3.264&  1.46024 &    2.67681&b3F-z3Go&          2&  \\
1.019600&       N {\scshape ii}&  -0.790&   4.08956&   5.30564&1Po-5/2[5/2]&       2& \\
1.019790&       Fe {\scshape i}&  -3.699&   2.72773&   3.94359&a3G-z3Fo &          2& \\
1.021910&       Fe {\scshape i}&  -0.215&   4.73346&   5.94680&y3Do-e3F &          2& *  \\
1.022120&       Fe {\scshape i}&  -2.760&   3.07155&   4.28464&c3P-z3Po &          2& \\
1.026800&       Fe {\scshape i}&  -4.467&   2.22286&   3.43042&a5P-z5Fo &         2B& \\
1.029180&       Si {\scshape i}&  -1.360&   4.92042&   6.12519& 3Po-3S  &         2B&\\
1.033010&      Sr {\scshape ii}&  -0.353&   1.83958&   3.03988& 2D-2Po  &          3& *  \\
1.033310&       N {\scshape ii}&  -1.228&   4.10569&   5.30564& 3Go-3F  &          2& \\
1.034370&       Fe {\scshape i}&  -3.493&   2.19801&   3.39674&a5P-z5Fo &          3& * As blend  \\
1.034670&       Ca {\scshape i}&  -0.408&   2.93271&   4.13109& 1Po-1S  &          3& * As blend, $\dag$\\
1.037411&        SiI & -0.447&  4.92965&   6.12519& 3Po-3S  &         2& *  \\
1.038150&       N {\scshape ii}&  -0.965&   4.08846&   5.28282&   ?     &         2B &\\
1.038184&        Fe {\scshape i}&  -4.065&  2.22286&   3.41718&a5P-z5Fo &         2B &\\
1.039860&       Fe {\scshape i}&  -3.330&   2.17609&   3.36848&a5P-z5Fo &         3B& * As blend  \\
1.039961&       Ti {\scshape i}&  -1.537&   0.84848&   2.04076&a5F-z5Go &         3B& * As blend  \\
1.042590&       Fe {\scshape i}&  -3.776&   2.69259&   3.88187&a3G-z3Fo &         2B& \\
1.042660&       Fe {\scshape i}&  -2.916&   3.07155&   4.26074&c3P-z3Po &         2B& *  \\
1.045544&     [Fe {\scshape i}]&     ?  &   0.99   &   2.17   &a3F-a5P  &  2& In Arcturus list   \\
1.045830&        S {\scshape i}&   0.260&   6.86061&   8.04620& 3So-3P  &         3F &\\
1.045960&        S {\scshape i}&  -0.430&   6.86061&   8.04605& 3So-3P  &         2F &\\
1.046230&        S {\scshape i}&   0.040&   6.86061&   8.04575& 3So-3P  &         3F &\\
1.047250&       Fe {\scshape i}&  -1.229&   3.88377&   5.06775&z3Do-3P  &          2& *  \\
1.048910&       Cr {\scshape i}&  -0.972&   3.01083&   4.19294&b5Dz5Do  &          2& \\
1.049900&       Ti {\scshape i}&  -1.651&   0.83605&   2.01705&a5F-z5Go &          3& \\
1.050438&      Fe {\scshape ii}&  -1.997&   5.54914&   6.72953&z4Fo-b4G &         2F& \\
1.051290&       Cr {\scshape i}&  -1.558&   3.01351&   4.19294&b5Dz5Do  &          2& \\
1.05324 &        P {\scshape i}&  0.200 &    6.95482 & 8.13207& 4P-4Do  &         2F& \\
1.053340&       N {\scshape ii}&  -1.189&   4.10569&   5.28282&   ??    &         2 & \\
1.053510&       Fe {\scshape i}&  -1.532&   3.92887&   5.10582&z3Do-3P  &          2& \\
1.054410&        C {\scshape i}&  -1.290&   8.53767&   9.71361& 1P-1Po  &         2F& \\
1.058000&       Fe {\scshape i}&  -3.268&   3.30163&   4.47357&b3H-z3Go &          2 &\\
1.058450&        P {\scshape i}&   0.480&   6.98570&   8.15716&4P-4Do   &         2F &\\
1.058750&       Ti {\scshape i}&  -1.777&   0.82591&   1.99703&a5F-z5Go &         3B& * As blend  \\
1.058800&       Si {\scshape i}&  -0.020&   4.95413&   6.12519& 3Po-3S  &         3B& * As blend, $\dag$\\
1.060630&       Si {\scshape i}&  -0.380&   4.92998&   6.09902& 3Po-3P  &          3& * As blend, $\dag$  \\
1.061060&       Ti {\scshape i}&  -2.676&   0.84848&   2.01705&a5F-z5Go &          3& * As blend  \\
1.06146 &       Fe {\scshape i}&  -0.336&   6.16964&     7.33777&n7Fo-9/2[13/2]&    2 &     \\
1.06196 &       Fe {\scshape i}&  -3.257&   3.26733&     4.43491&b3H-z3Go&         2 & \\  
1.063060&      Si {\scshape i}]&   0.000&   5.86287&   7.02925& 1P-3Po  &          2 &\\
1.063883&      Al {\scshape ii}&  -0.610&  15.30358&  16.46905& 3Fo-3D  &     2F&\\
1.065050&       Cr {\scshape i}&  -1.613&   3.01083&   4.17502&b5Dz5Do  &          2& \\
1.066390&       Si {\scshape i}&  -0.350&   4.92042&   6.08315& 3Po-3P  &         3B& * As blend, $\dag$  \\
1.066460&       Ti {\scshape i}&  -1.916&   0.81820&   1.98085&a5F-z5Go &         3B& * As blend  \\
1.067047&       Cr {\scshape i}&  -1.489&   3.01301&   4.17502&b5Dz5Do  &          2& \\
1.067510&       Cr {\scshape i}&  -1.374&   3.01351&   4.17502&b5Dz5Do  &          2& \\
1.068000&       Ti {\scshape i}&  -2.507&   0.83605&   1.99703&a5F-z5Go &          3& \\
1.068600&        C {\scshape i}&   0.076&   7.48328&   8.64360& 3Po-3D  &         3F& \\
1.068830&        C {\scshape i}&  -0.276&   7.48090&   8.64098& 3Po-3D  &         3F& \\
1.069260&       Si {\scshape i}&  -0.190&   5.95413&   7.11374& 3D-3Fo  &          2& *  \\
1.069420&        C {\scshape i}&   0.348&   7.48830&   8.64774& 3Po-3D  &         3F& \\
1.069720&       Si {\scshape i}&  -0.060&   5.96433&   7.12345& 3D-3Fo  &          2& \\
1.071030&        C {\scshape i}&  -0.401&   7.48328&   8.64098& 3Po-3D  &         3F& \\
1.072920&       Ti {\scshape i}&  -2.064&   0.81300&   1.98652&a5F-z5Go &         3B& \\
1.073030&       Si {\scshape i}&   0.060&   5.98441&   7.13995& 3D-3Fo  &         3B& \\
1.073250&        C {\scshape i}&  -0.401&   7.48830&   8.64360& 3Po-3D  &         3F& \\
1.073590&       Ti {\scshape i}&  -2.504&   0.82591&   1.98085&a5F-z5Go &          3& \\
1.075230&       Si {\scshape i}&  -0.300&   4.92998&   6.08315& 3Po-3P  &          3& \\
1.075595&       Fe {\scshape i}&  -1.902&   3.95999&   5.11277& z3Do-3P &          2& \\
1.075690&        C {\scshape i}&  -1.590&   7.48830&   8.64098& 3Po-3D  &         2F& \\
1.077770&       Ti {\scshape i}&  -2.658&   0.81814&   1.96852&a5F-z5Go &          3& \\
1.07836 &       Fe {\scshape i}&  -3.414&   3.23694&   4.38676& b3H-z3Go&          2& \\  
1.078600&       Fe {\scshape i}&  -2.597&   3.11117&   4.26074&c3P-z3Po &          2& \\
1.078751&       Si {\scshape i}&  -0.910&   5.96433&   7.11374& 3D-3Fo  &          2& \\  
1.078980&       Si {\scshape i}&  -0.390&   4.92998&   6.07914& 3Po-3P  &          3& \\
1.08043 &       Cr {\scshape i}&  -1.715&   3.01135&   4.15897& b5D-z5Do&   2    &\\
1.081410&       Mg {\scshape i}&  -0.320&   5.94632&   7.09290& 3D-3Fo  &         2B& $\dag$  \\
1.082130&       Fe {\scshape i}&  -1.998&   3.95999&   5.10582&z3Do-3P  &         2B &\\
1.082470&       Cr {\scshape i}&  -1.678&   3.01351&   4.15897&b5D-z5Do &          2 &\\
1.083010&       Si {\scshape i}&   0.220&   4.95413&   6.09902& 3Po-3P  &          3 &\\
1.083091&       Ti {\scshape i}&  -3.878&   0.83605&   1.98085&a5F-z5Go &         2B&\\
1.08378&        Na {\scshape i}&  -0.500&   3.61721&   4.76129&2D-2Fo   &       2& $\dag$   \\
1.084680&       Si {\scshape i}&  -0.310&   5.86287&   7.00599& 1P-1Do  &          2 &\\
1.086562&      Fe {\scshape ii}&  -2.121&   5.58957&   6.73071&z4Fo-b4G &         2F &\\
1.086650&       Fe {\scshape i}]&  -0.803&   4.73346&   5.87452&y3Do-e5F &          2& \\
1.087180&       Si {\scshape i}&   0.040&   6.19142&   7.33192&3Fo-3/2[9/2]       &  2B& \\
1.087250&       Si {\scshape i}&   0.320&   5.08269&   6.22311& 1Po-1D  &         3B& $\dag$  \\
1.08847 &       Fe {\scshape i}]&  -3.713&   2.84523&   3.98438& b3P-z3Fo   &         2&\\
1.088578&       Si {\scshape i}&  -0.870&   5.98441&   7.12345& 3D-3Fo  &          2&\\
1.08883 &       Si {\scshape i}&   0.040&   6.18113&   7.31990&3Fo-3/2[7/2] &         2 & \\
1.089930&       Fe {\scshape i}&  -2.729&   3.07155&   4.20917&c3P-z3Po &          2 &\\
1.090886&       Cr {\scshape i}&  -0.647&   3.43820&   4.57480&y7Po-e7S &          2 &\\
1.091723&      Mg {\scshape ii}&   0.020&   8.86425&  10.00002&  2D-2Po  &  2FB&\\
1.091790&      Sr {\scshape ii}&  -0.638&   1.80482&   2.94051& 2D-2Po  &   3 &\\
1.091827&      Mg {\scshape ii}&  -0.930&   8.86436&  10.00002&  2D-2Po  &  2FB&\\
1.093287&        Cr {\scshape i}& -2.330&    3.01351&  4.15897& b5D-z5Do &    2& Amidst CN bands \\
1.094110&        H {\scshape i}&   0.002&  12.08830&  13.22160&  3-6    &         3F &\\
1.095477&      Mg {\scshape ii}&  -0.230&   8.86436&  9.99622&  2D-2Po  &  2F &   \\
1.095630&       Mg {\scshape i}&  -0.852&   5.93194&   7.06364& 3Po-3D  &         2B& $\dag$  \\
1.096023&        Cr {\scshape i}& -2.067&    3.01351&  4.15897& b5D-z5Do &   2B &\\
1.096030&       Mg {\scshape i}&  -0.690&   5.93235&   7.06364& 3Po-3D  &    2B & $\dag$ \\
1.096845&       Mg {\scshape i}&  -0.250&   5.93319&   7.06363& 3Po-3D  & 2B& CN band head confused, $\dag$  \\
1.097038&        P {\scshape i}&   0.120&   8.07889&   9.20914& 2D-2Do  &     2F& \\
1.098231&       SiI & -0.590&   4.95413&   6.08315 & 3Po-3P  &         3& \\
1.098284&       N {\scshape ii}&  -0.929&   4.15386&   5.28282&y3Do-e3D &        3F &\\
1.09851 &       Si {\scshape i}&  -0.050&   6.19142&   7.32015&3Fo-3/2[7/2]& 2& \\
1.09875 &       Si {\scshape i}&  -0.500&   6.19142&   7.31990&3Fo-3/2[7/2]& 2& \\
1.10163&        Fe {\scshape i}]&  -1.334&  4.79579&    5.92133&y3Do-e5F &    2B &\\ 
1.101870&       Cr {\scshape i}&  -0.515&   3.43820&   4.57480&y7Po-e7S &          2 &\\
1.102100&       Si {\scshape i}&   0.610&   6.20642&   7.33148&3Fo-3/2[9/2]     &   3& \\
1.102287&       KI &  -0.010&   2.67014&   3.79501 &  2D-2Fo &     2& Doublet, wavelength off, $\dag$ \\
1.104770&       Cr {\scshape i}&  -2.508&   3.01135&   4.13369&b5D-z5Do &          2 & \\
\enddata
\tablecomments{
This subset of the lines in our assembled line list includes those lines seen among the standards grid that are rated at 3 or 2 in strength (i.\/e.\/, absorption below 90\% of the continuum level).  Not tabulated are lines from our list rated 1 (weaker absorption than 90\% of the continuum) or 0 (not present at our spectral resolution or a likely misidentification) in strength.  Wavelengths are those in vacuum. In addition to the line strength, the ``Strength" column also indicates which lines are blended at greater than the 50\% level (marked with a B) and which  lines that are only seen in the F spectroscopic standards (marked with an F). The ``Comments" column indicates lines seen in the lower resolution $R=2000$ $Y$ band data of \citet{Cush05} (marked with an asterisk), lines identified as possibly present in the $R=250$ data of \citet{Leg96} (marked with a dagger), or other information.}
\end{deluxetable}

\pagebreak
\begin{deluxetable}{cccl}
\setlength{\tabcolsep}{0.3in}
\tablewidth{0pt}
\tablecaption{Sample Normalized Spectrum of HD 87, Order 69 \label{tab:asciiex}}
\tablehead{
{(1)}        &{(2)} \\
{$\lambda$ [$\mu$m]}  &{Normalized Flux}} 
\startdata
1.0963141 & 1.07008 \\
1.0963299 & 1.07359 \\
1.0963458 & 1.10119 \\
1.0963615 & 1.08921 \\
1.0963775 & 1.11744 \\
1.0963933 & 1.14190 \\
1.0964092 & 1.08628 \\
1.0964249 & 1.06581 \\
1.0964408 & 1.08024 \\
1.0964566 & 1.03010 \\
\enddata
\tablecomments{In this table, we present a sample demonstrating the form and content of the spectra available for download with the online edition of the journal. Similar spectra are available in the online journal (spectra.tar.gz).}
\end{deluxetable}

\clearpage

\begin{figure}\begin{center}
\includegraphics[height=7in, clip=false]{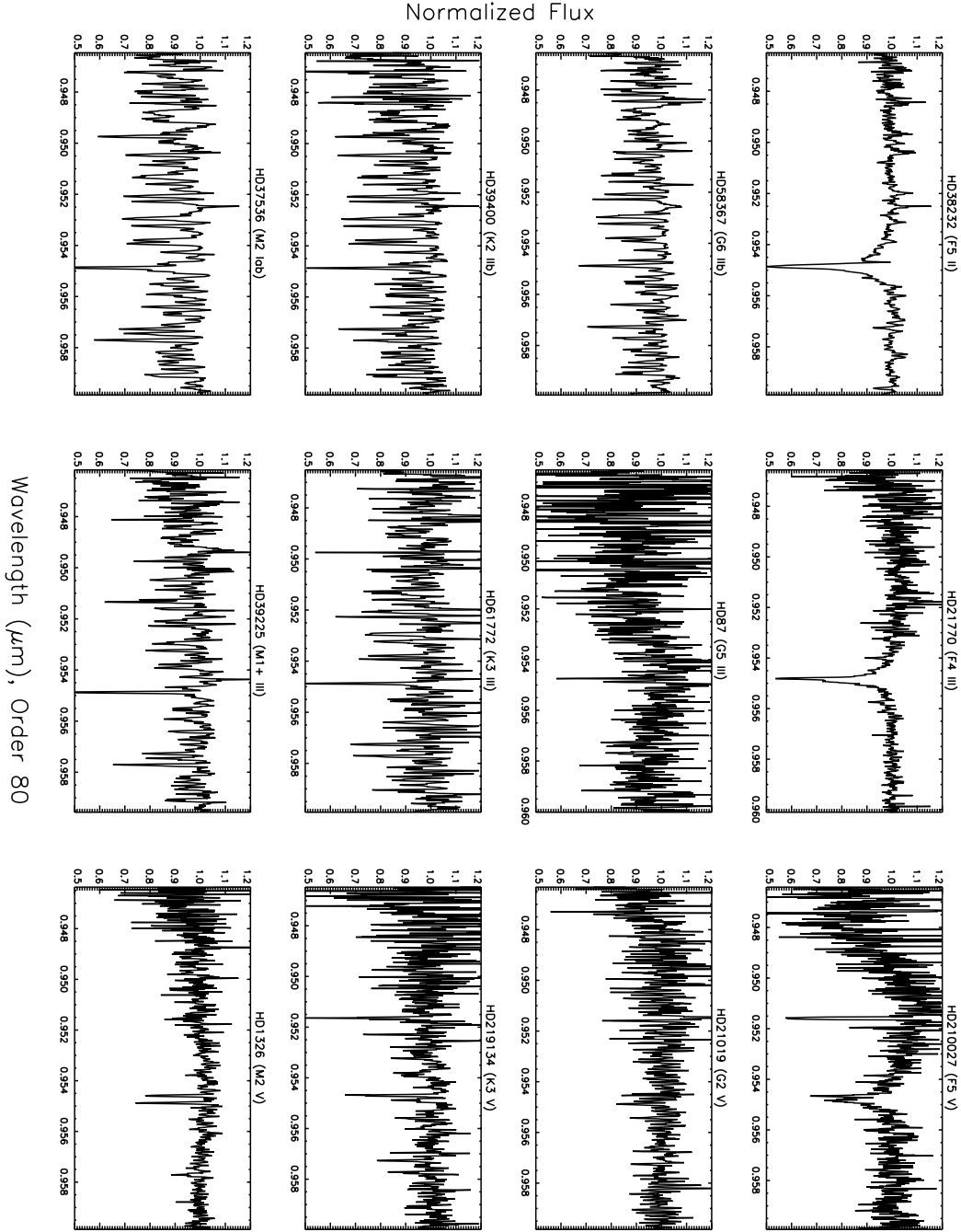}\end{center}
\figurenum{1a}
\caption{ Order 80 for the grid of spectroscopic standards. The four rows from top to bottom are the F, G, K, and M standards, and the three columns from left to right are the supergiants (I--II), giants (III), and dwarfs (IV--V).  Each NIRSPEC spectral order is represented in a subsequent figure. (b) Same as (a), order 79. (c) Same as (a), order 78. (d) Same as (a), order 77. (e) Same as (a), order 76. (f) Same as (a), order 75. (g) Same as (a), order 74. (h) Same as (a), order 73. (i) Same as (a), order 72. (j) Same as (a), order 71. (k) Same as (a), order 70. (l) Same as (a), order 69. \label{fig:grid_80}}
\end{figure}

\begin{figure}\begin{center}
\includegraphics[height=8in, clip=false]{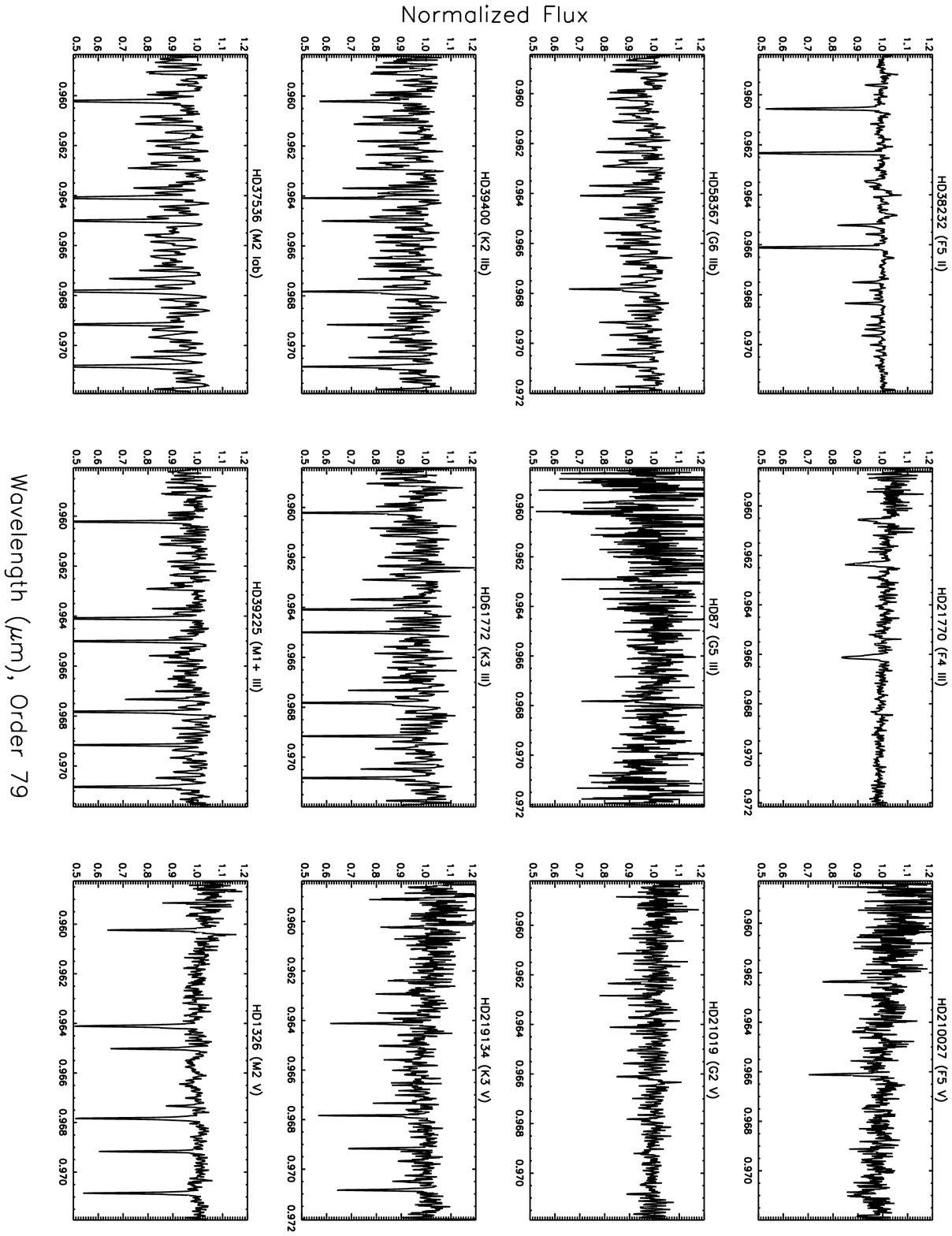}\end{center}
\figurenum{1b}
\caption{(Continued) \label{fig:grid_79}}
\end{figure}

\begin{figure}\begin{center}
\includegraphics[height=8in, clip=false]{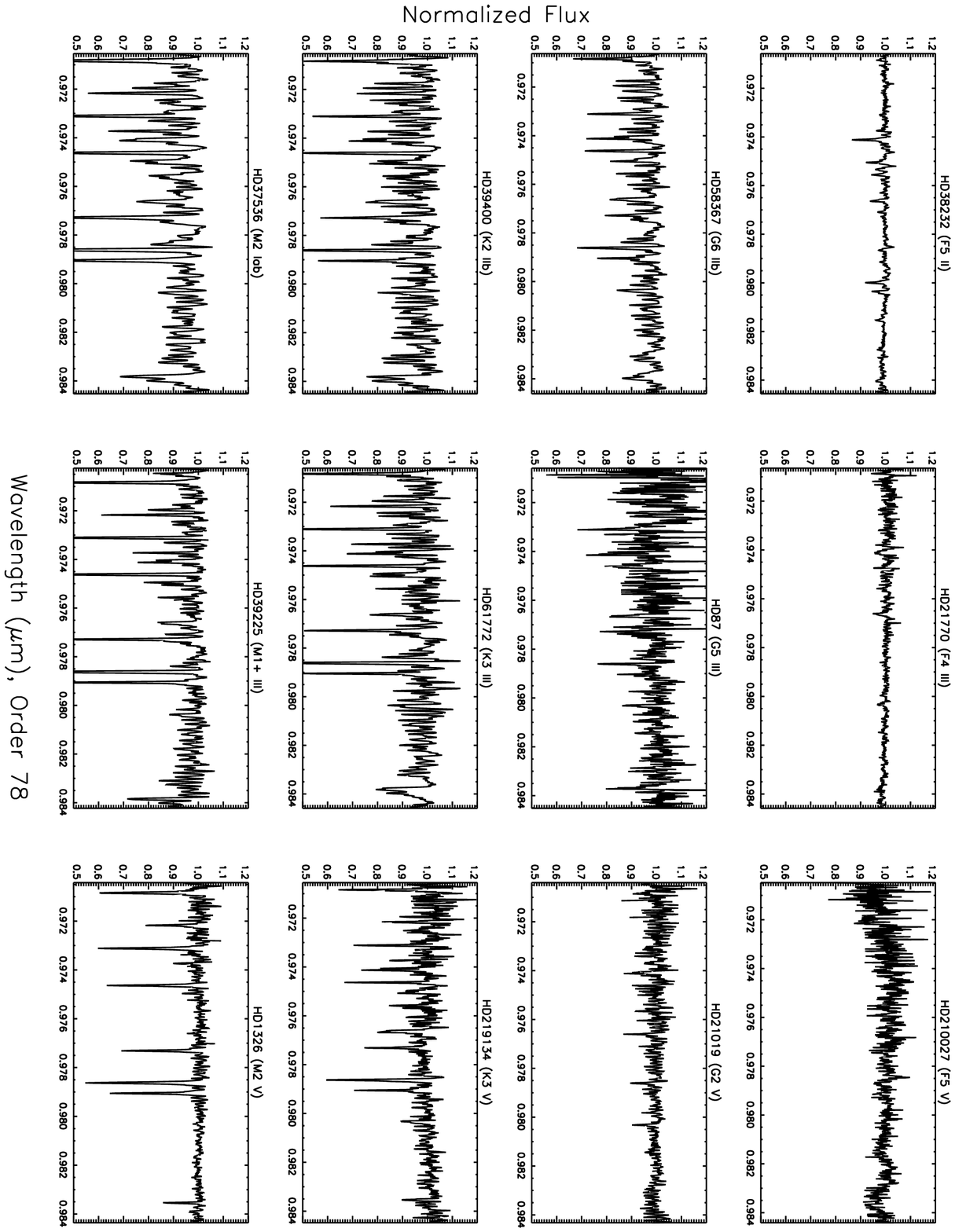}\end{center}
\figurenum{1c}
\caption{(Continued) \label{fig:grid_78}}
\end{figure}

\begin{figure}\begin{center}
\includegraphics[height=8in, clip=false]{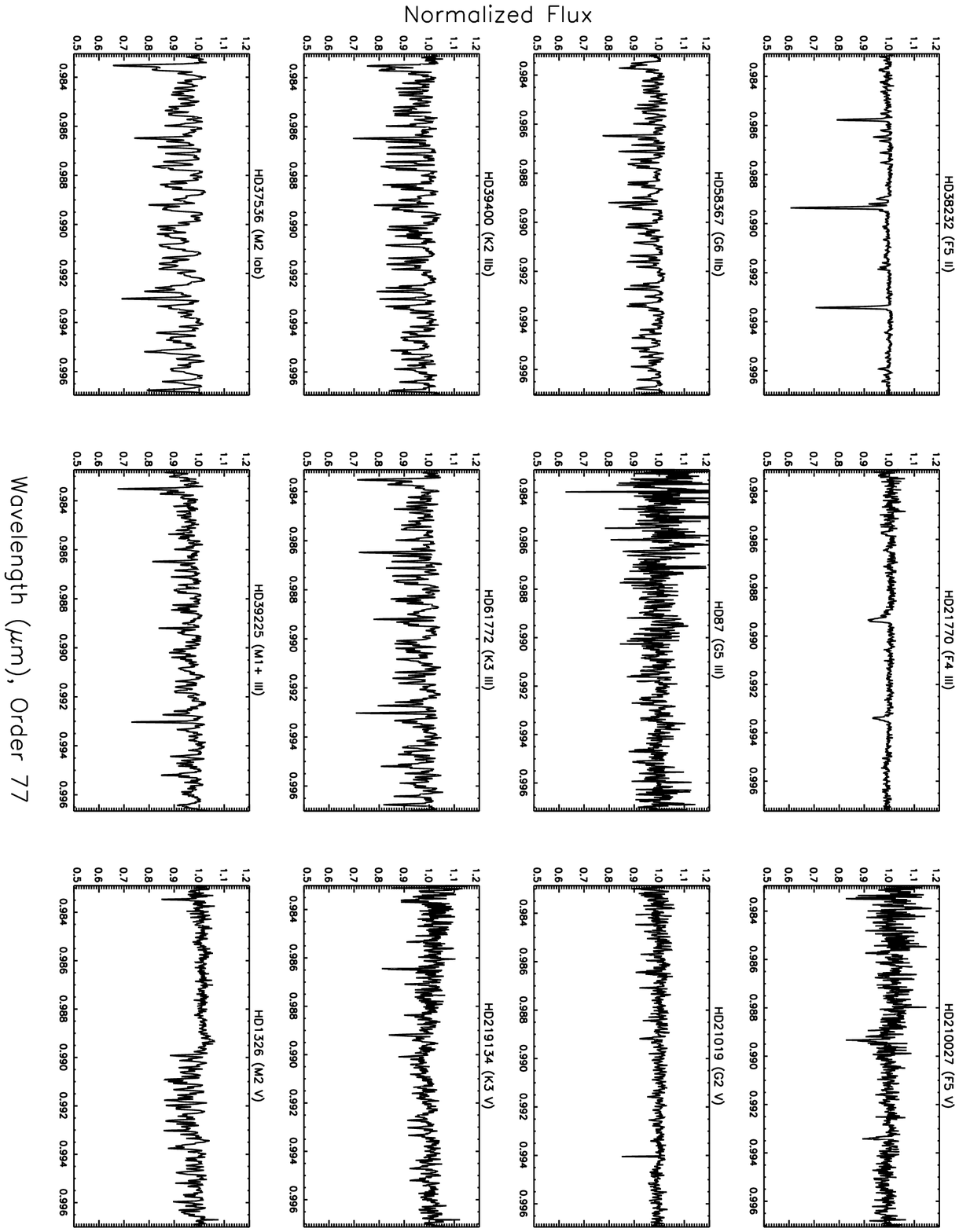}\end{center}
\figurenum{1d}
\caption{(Continued) \label{fig:grid_77}}
\end{figure}

\begin{figure}\begin{center}
\includegraphics[height=8in, clip=false]{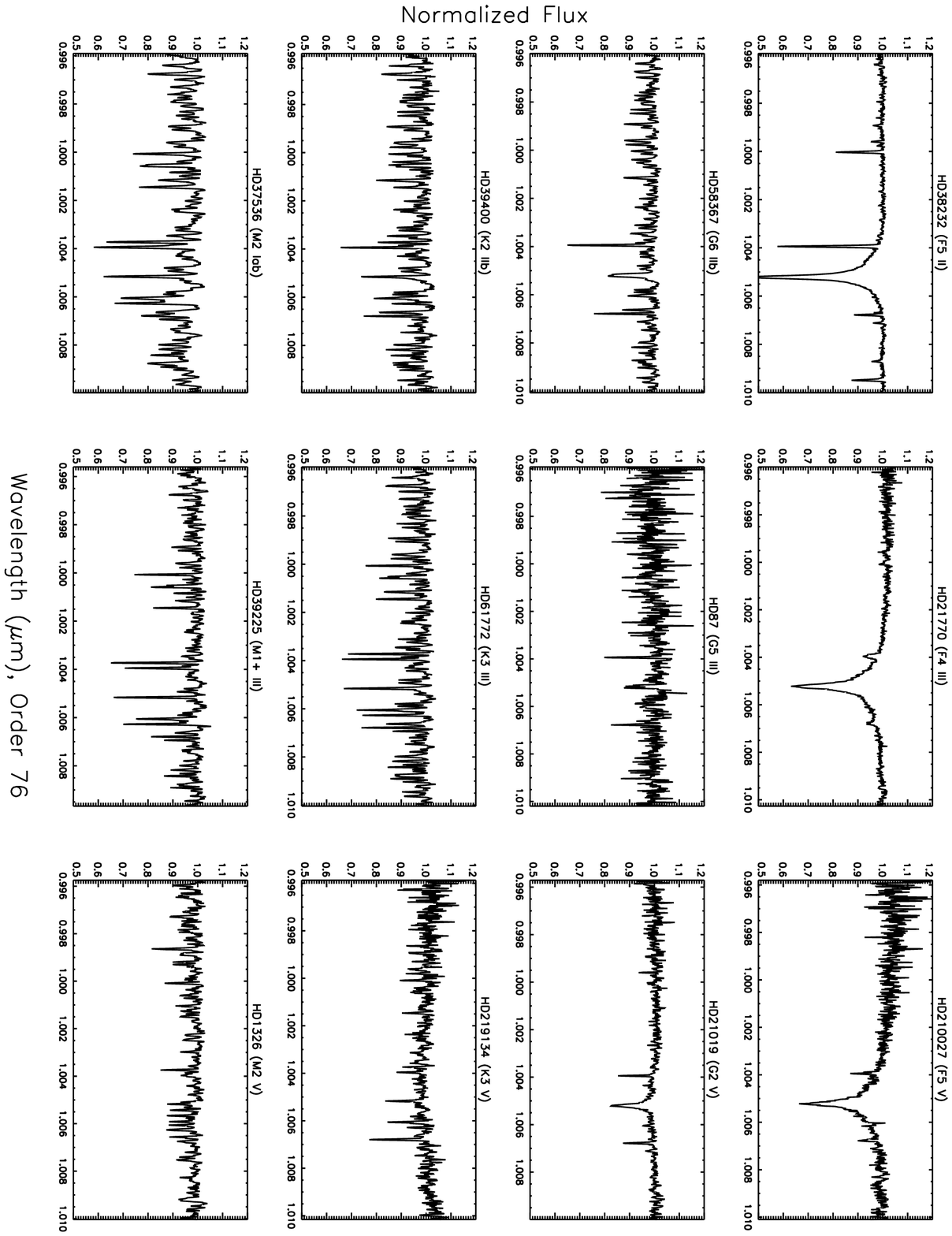}\end{center}
\figurenum{1e}
\caption{(Continued) \label{fig:grid_76}}
\end{figure}

\begin{figure}\begin{center}
\includegraphics[height=8in, clip=false]{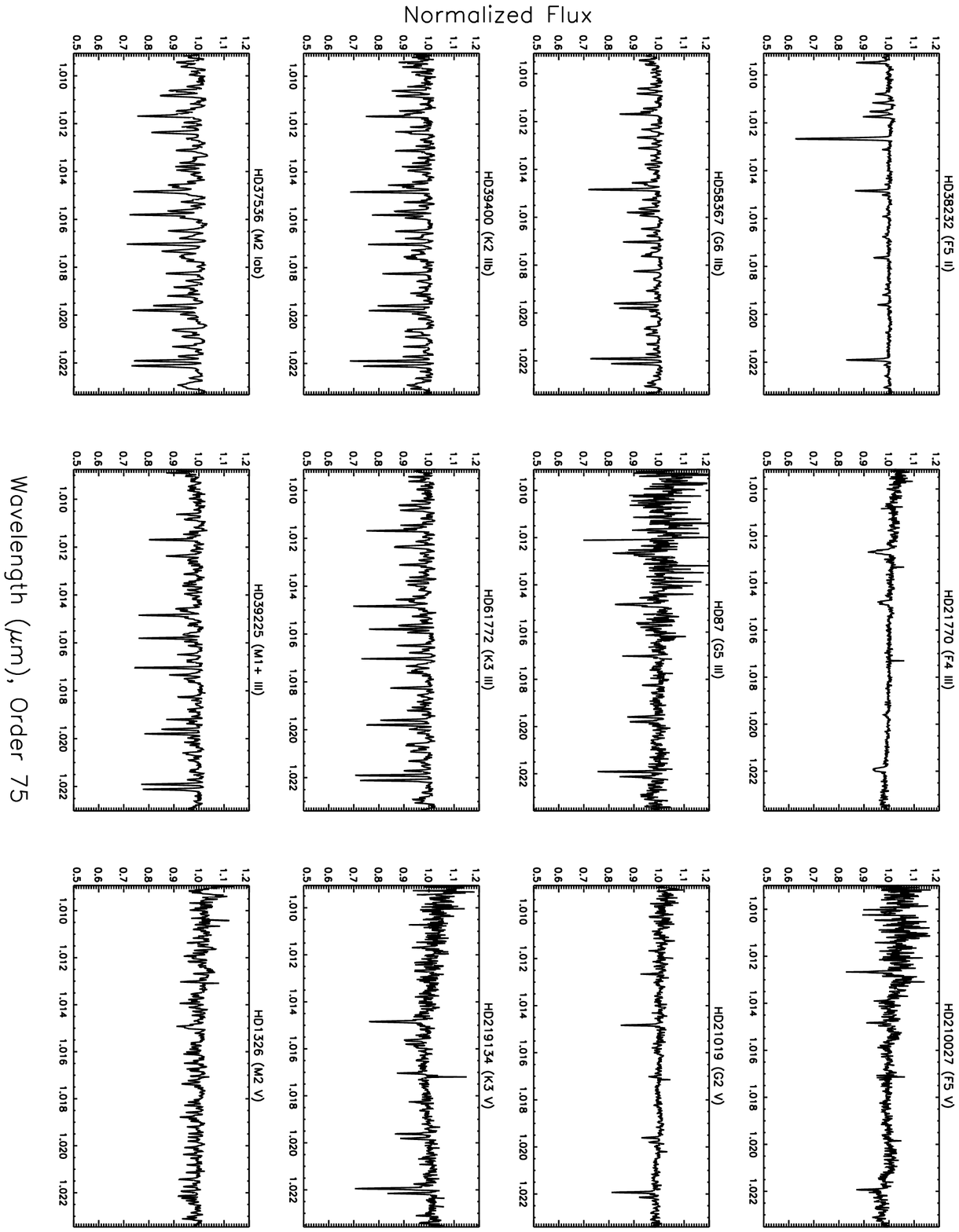}\end{center}
\figurenum{1f}
\caption{(Continued) \label{fig:grid_75}}
\end{figure}

\begin{figure}\begin{center}
\includegraphics[height=8in, clip=false]{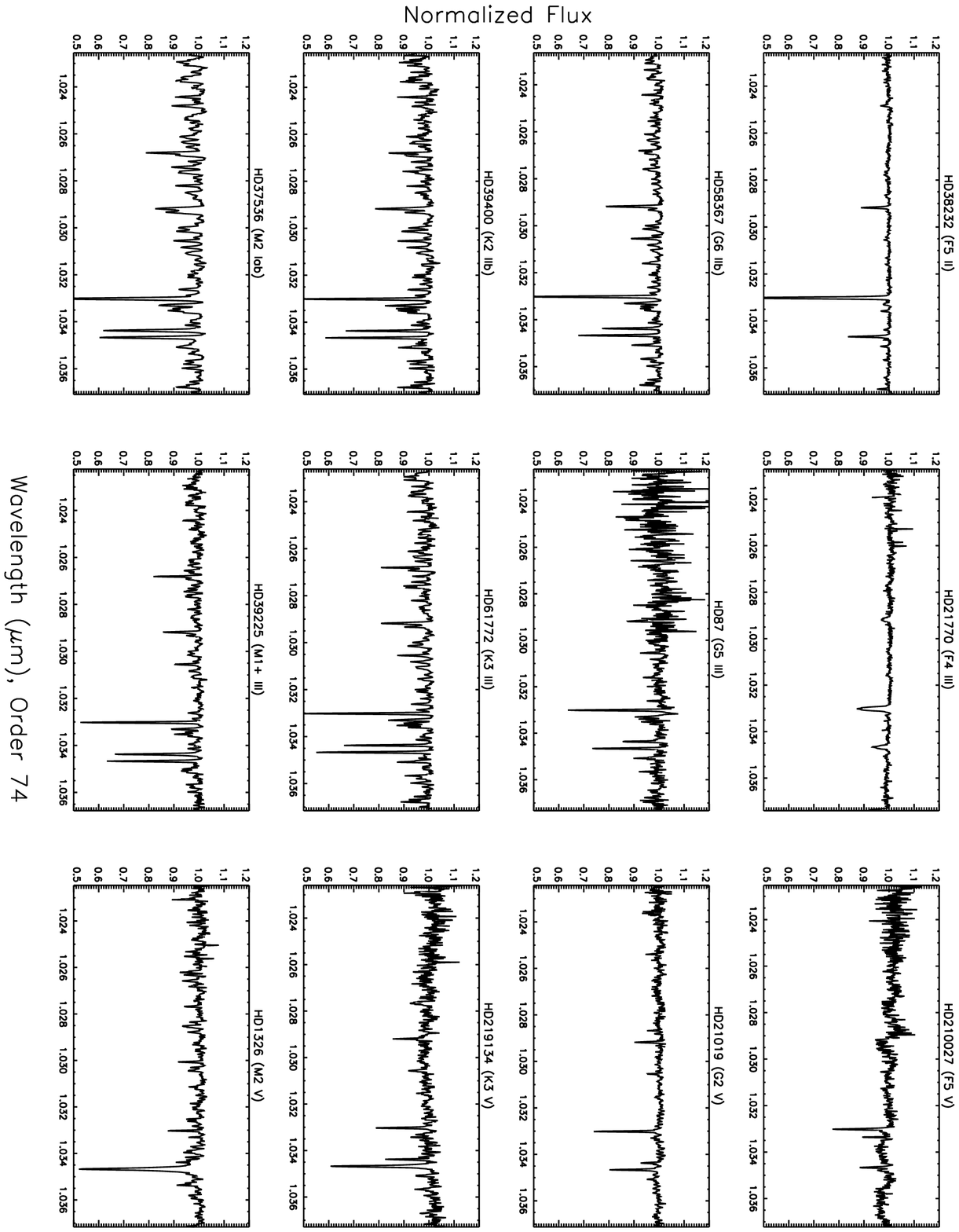}\end{center}
\figurenum{1g}
\caption{(Continued) \label{fig:grid_74}}
\end{figure}

\begin{figure}\begin{center}
\includegraphics[height=8in, clip=false]{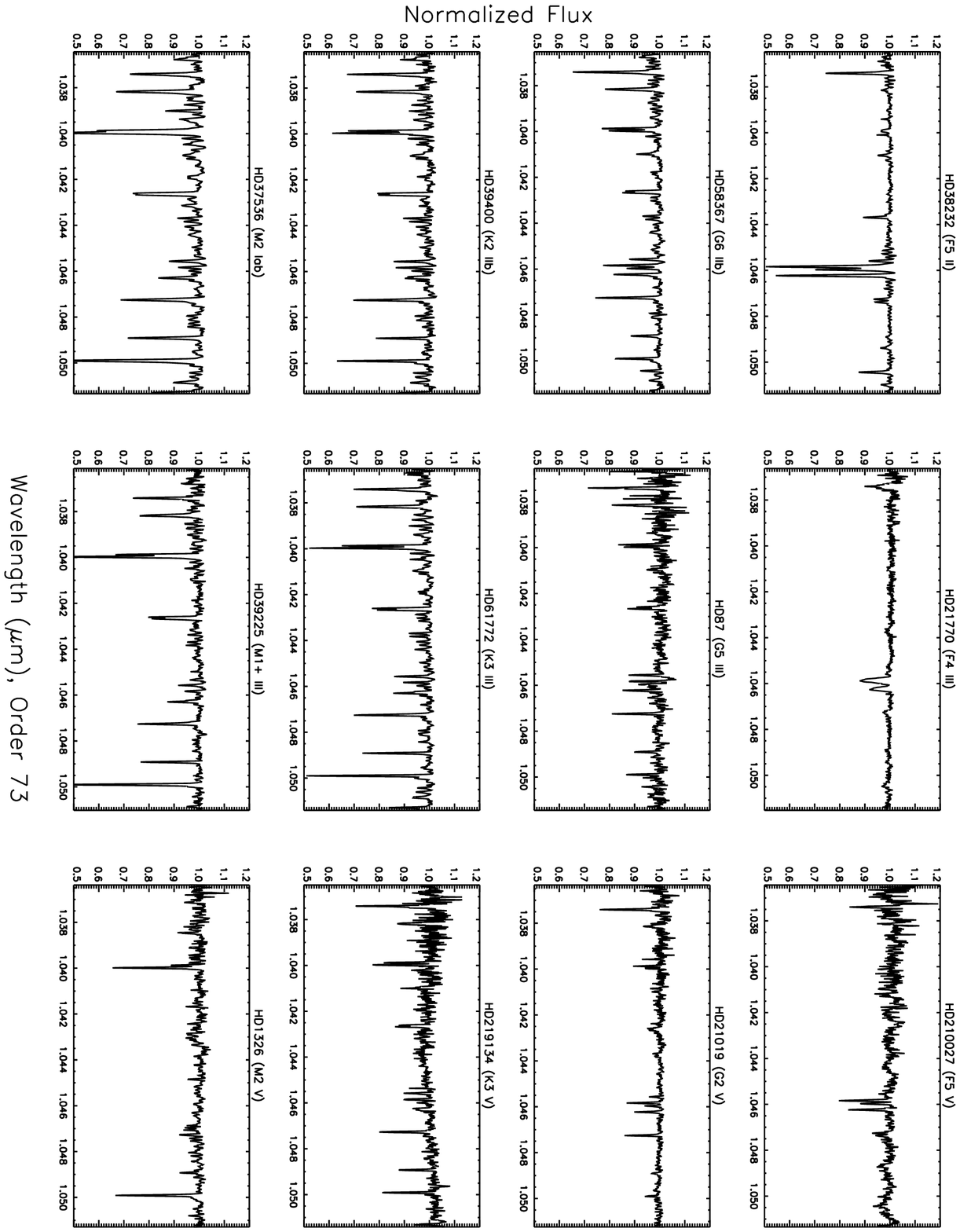}\end{center}
\figurenum{1h}
\caption{(Continued) \label{fig:grid_73}}
\end{figure}

\begin{figure}\begin{center}
\includegraphics[height=8in, clip=false]{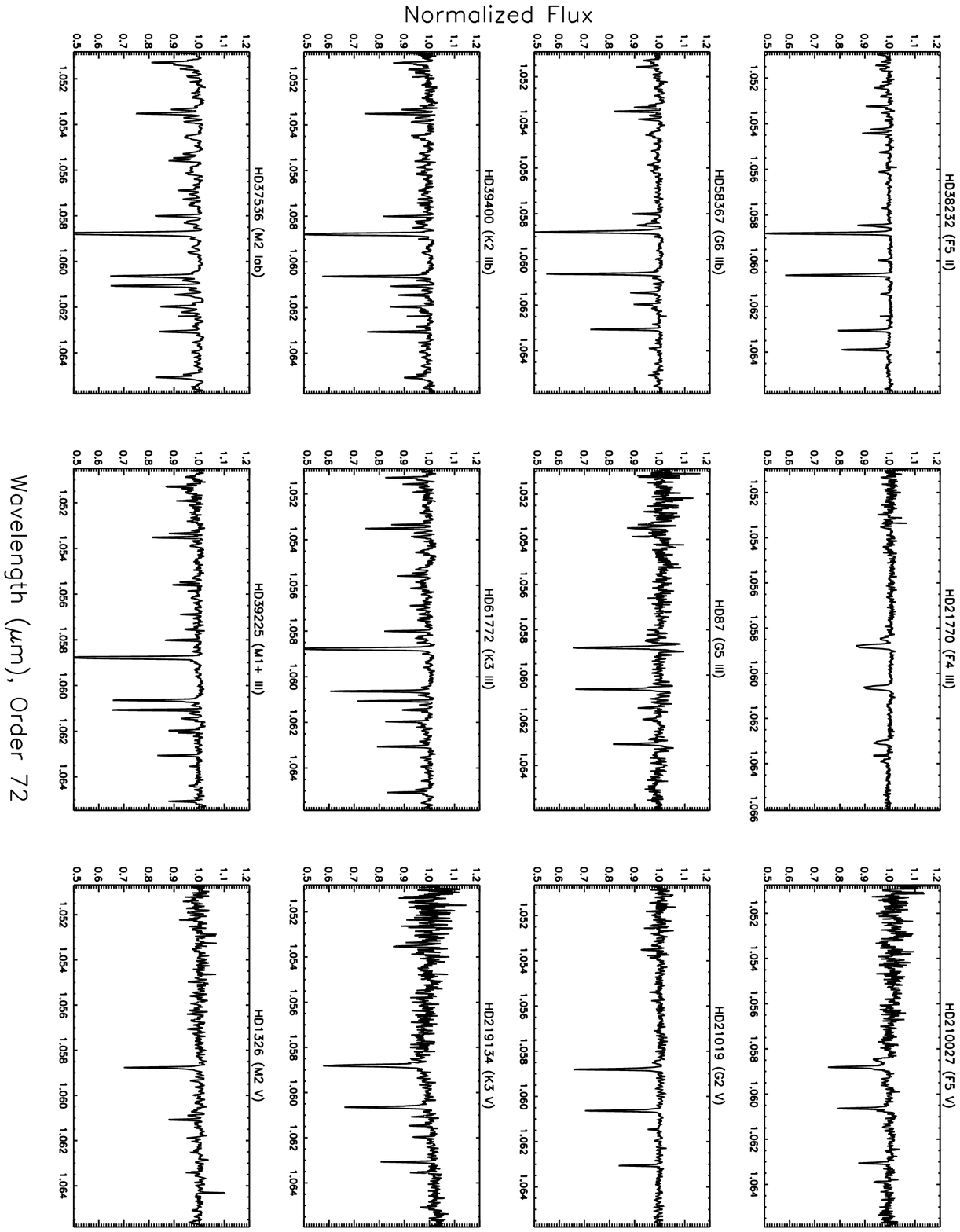}\end{center}
\figurenum{1i}
\caption{(Continued) \label{fig:grid_72}}
\end{figure}

\begin{figure}\begin{center}
\includegraphics[height=8in, clip=false]{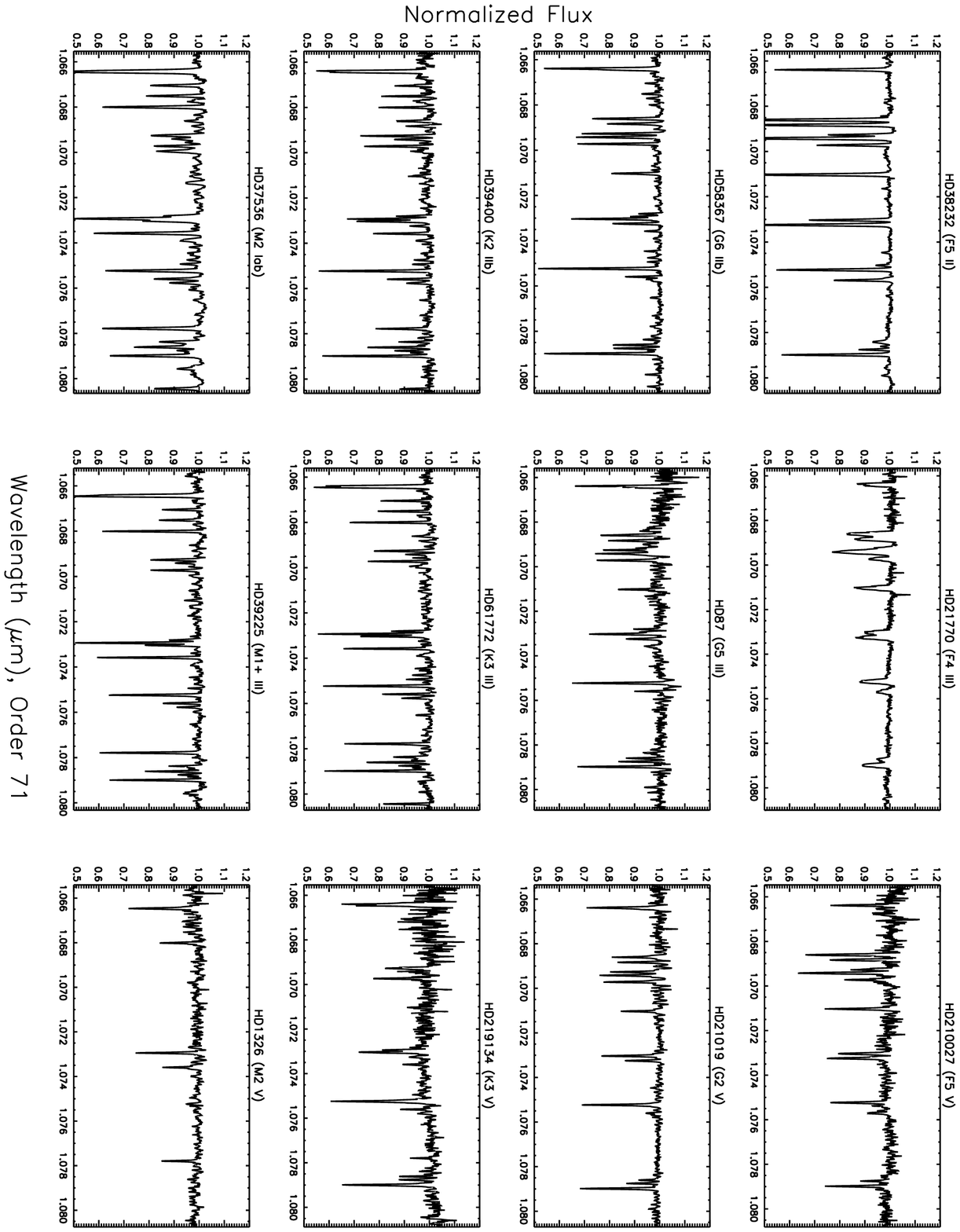}\end{center}
\figurenum{1j}
\caption{(Continued) \label{fig:grid_71}}
\end{figure}

\begin{figure}\begin{center}
\includegraphics[height=8in, clip=false]{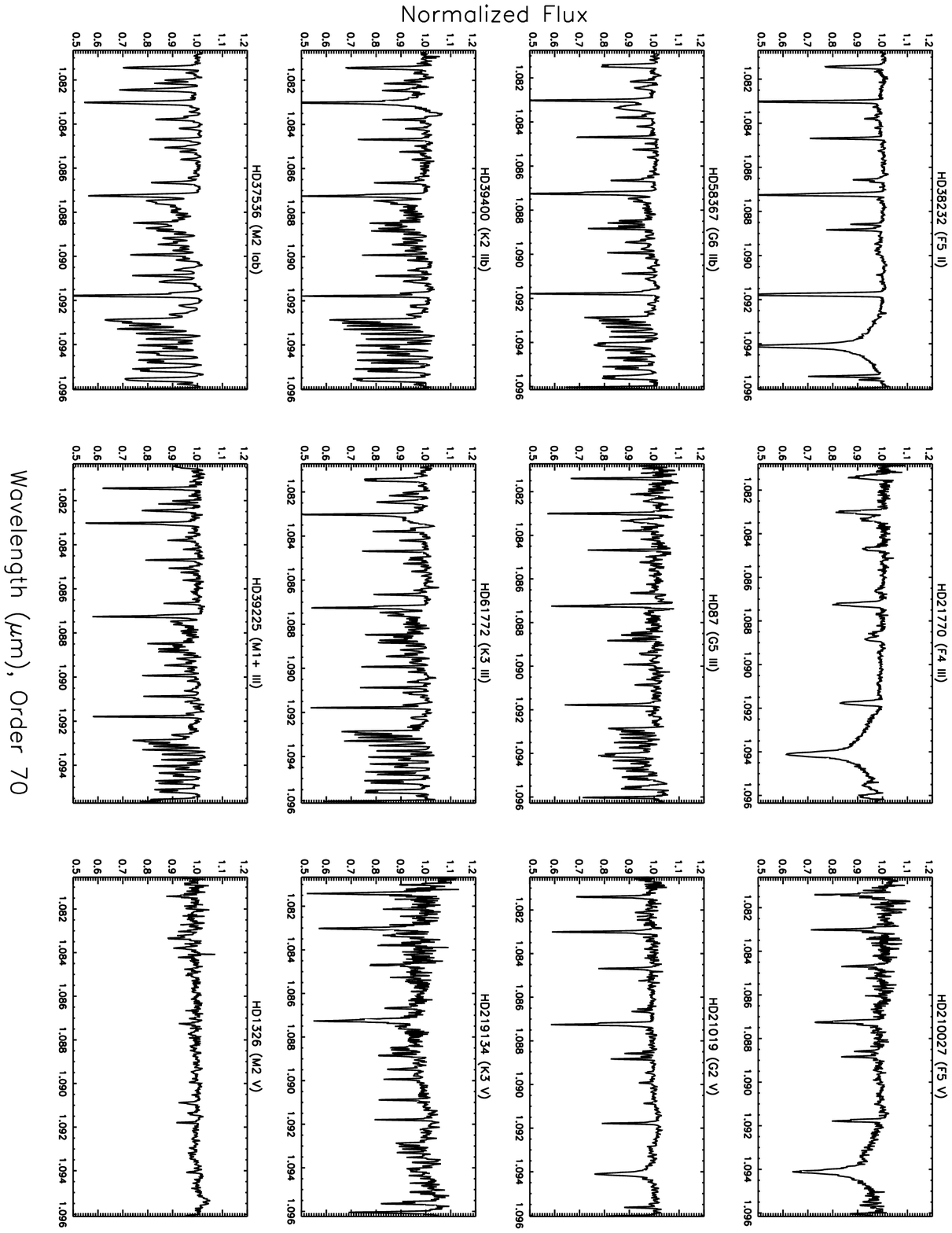}\end{center}
\figurenum{1k}
\caption{(Continued) \label{fig:grid_70}}
\end{figure}

\begin{figure}\begin{center}
\includegraphics[height=8in, clip=false]{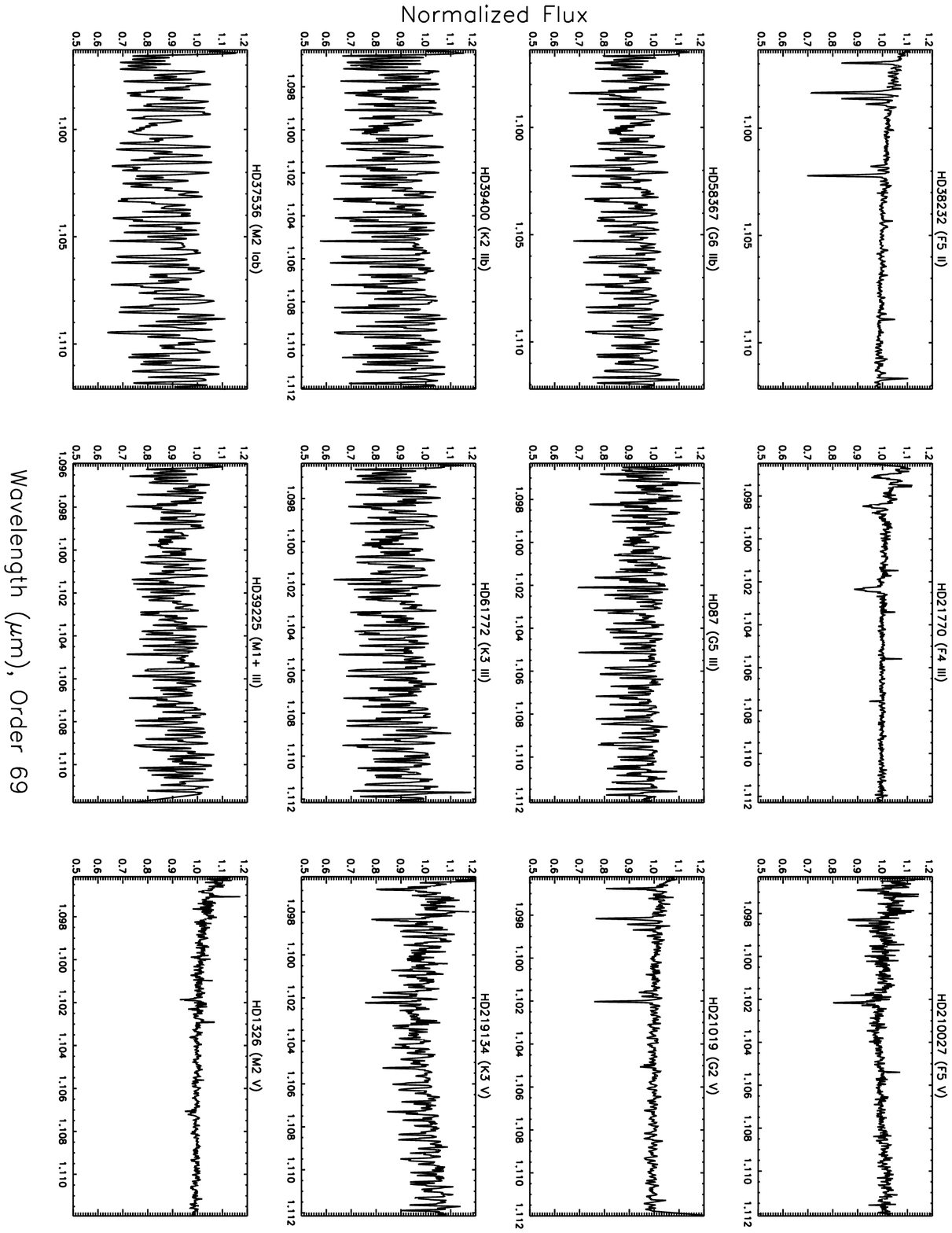}\end{center}
\figurenum{1l}
\caption{(Continued) \label{fig:grid_69}}
\end{figure}

\clearpage

\clearpage

\begin{figure}\begin{center}
\includegraphics[height=7in, clip=true]{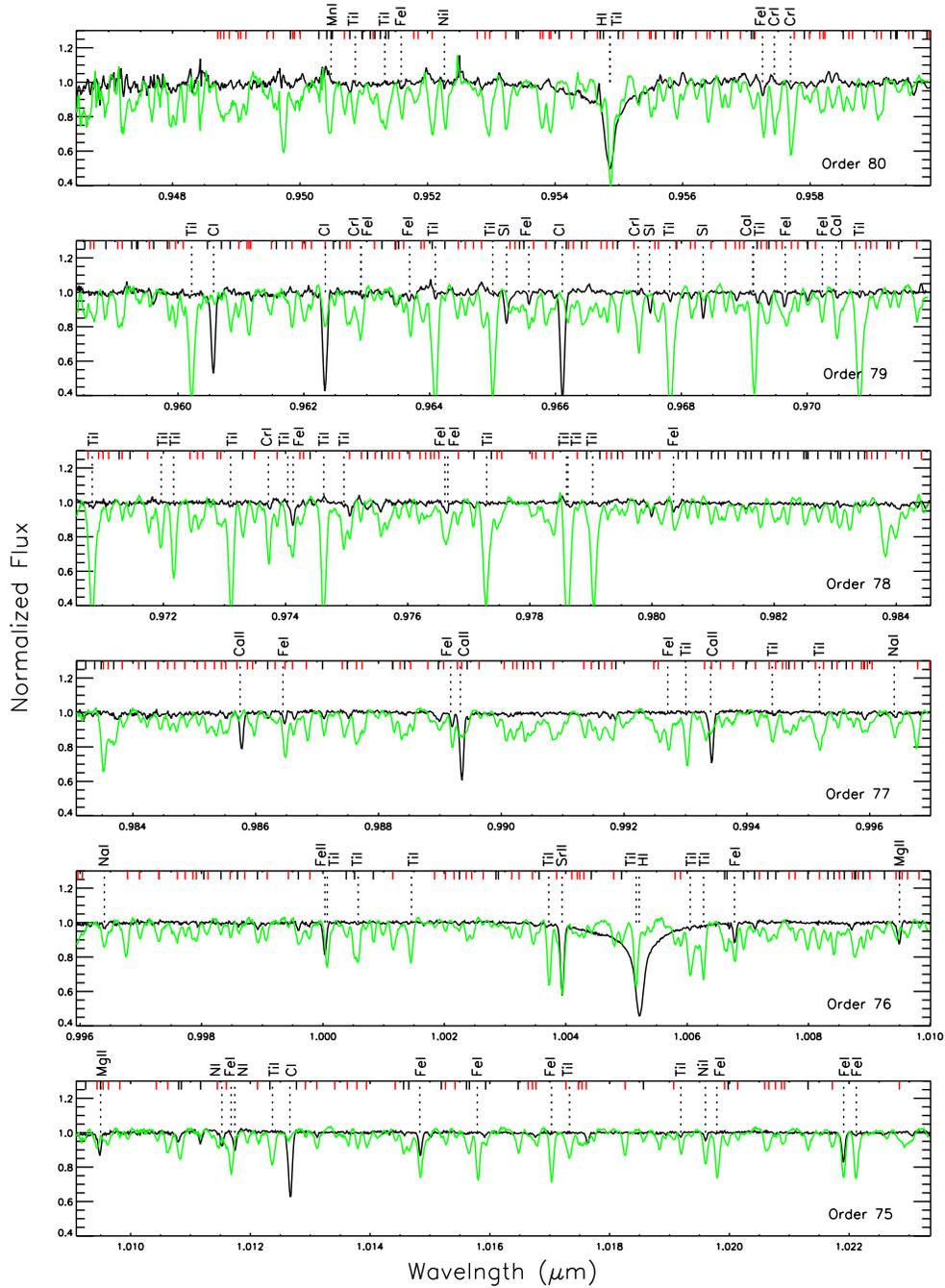}\end{center}
\figurenum{4a}
\caption{Strong atomic lines tabulated in Table ~\ref{tab:lines} (vertical dotted lines with element names) identified relative to the supergiant spectroscopic standards HD 38232, an F5 {\scshape ii} (black line) and HD 37536, an M2 Iab (green line). Weaker atomic lines from our completed catalog are marked at the top (short black hashes) as are molecular lines (short red hashes) (A color version of this figure is available in the online journal.).
\label{fig:fmgiant_pg1}}
\end{figure}

\begin{figure}\begin{center}
\includegraphics[height=7in, clip=true]{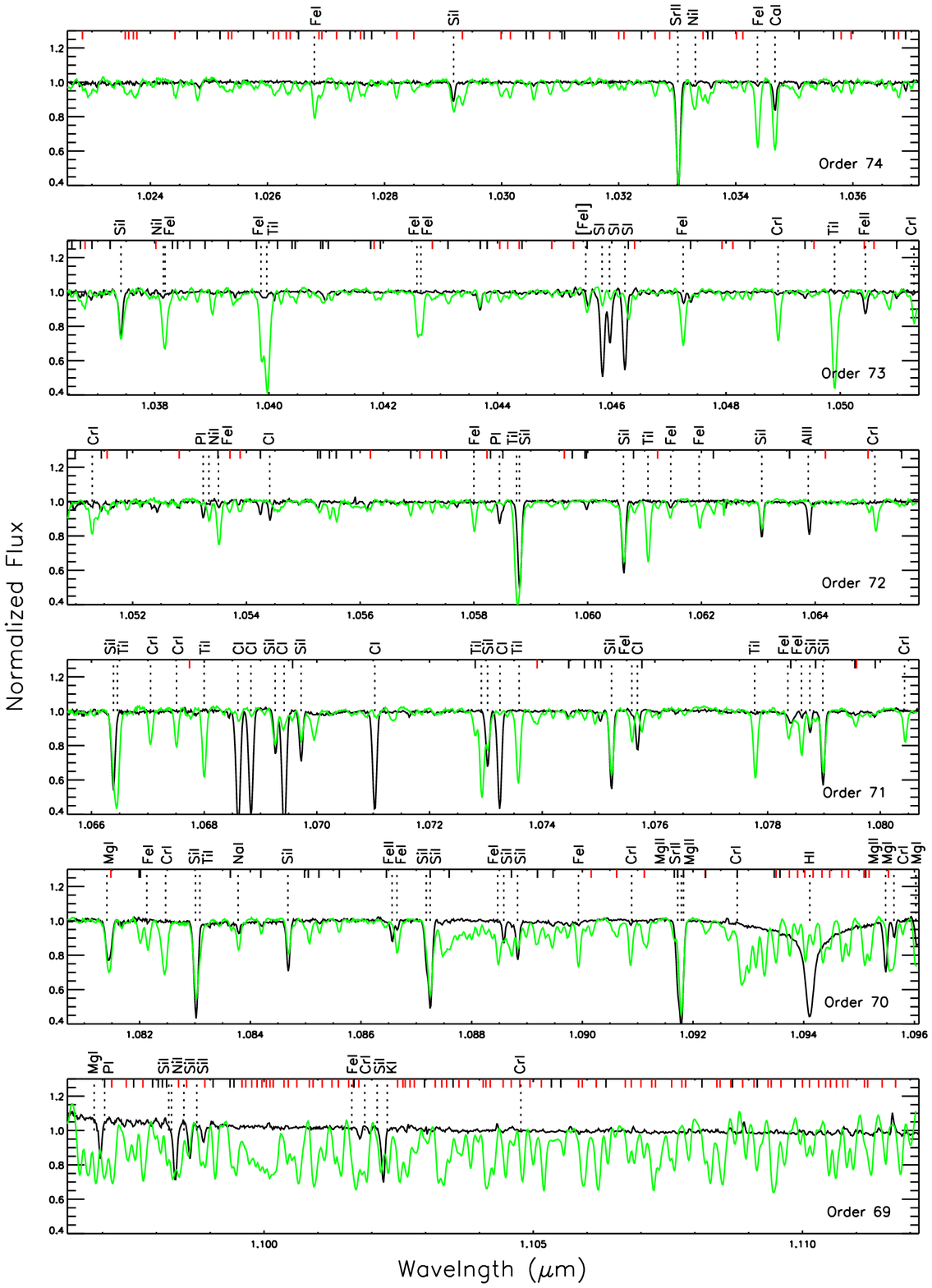}\end{center}
\figurenum{4b}
\caption{(Continued) \label{fig:fmgiant_pg2}}
\end{figure}

\begin{figure}\begin{center}
\includegraphics[height=7in, clip=true]{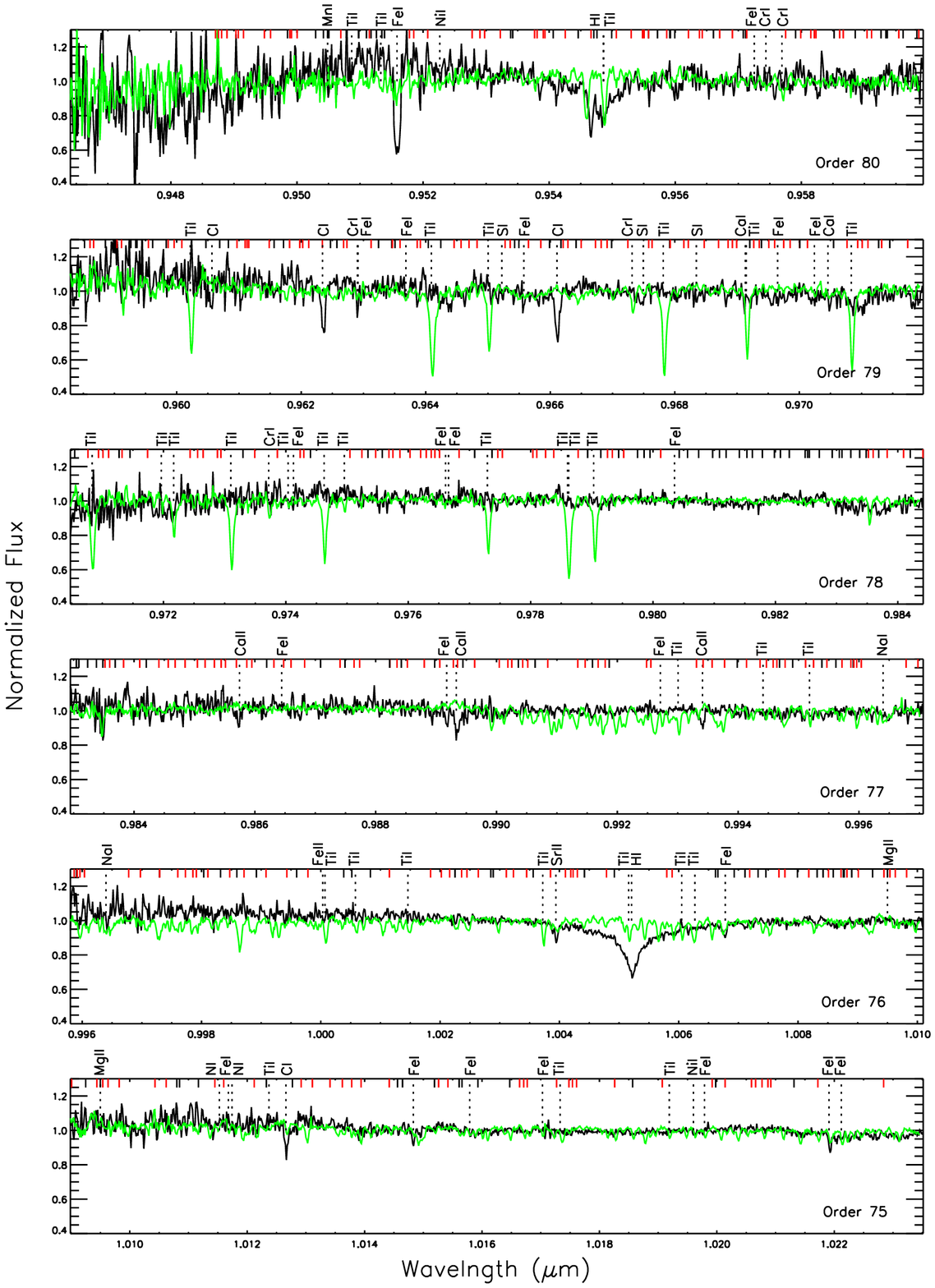}\end{center}
\figurenum{5a}
\caption{Strong atomic lines tabulated in Table ~\ref{tab:lines} (vertical dotted lines with element names) identified relative to the dwarf spectroscopic standards HD 55052, an F5 V (black line) and HD 1326, an M2 V (green line). Weaker atomic lines from our completed catalog are marked at the top (short black hashes) as are molecular lines (short red hashes) (A color version of this figure is available in the online journal.). 
\label{fig:fmdwarf_pg1}}
\end{figure}

\begin{figure}\begin{center}
\includegraphics[height=7in, clip=true]{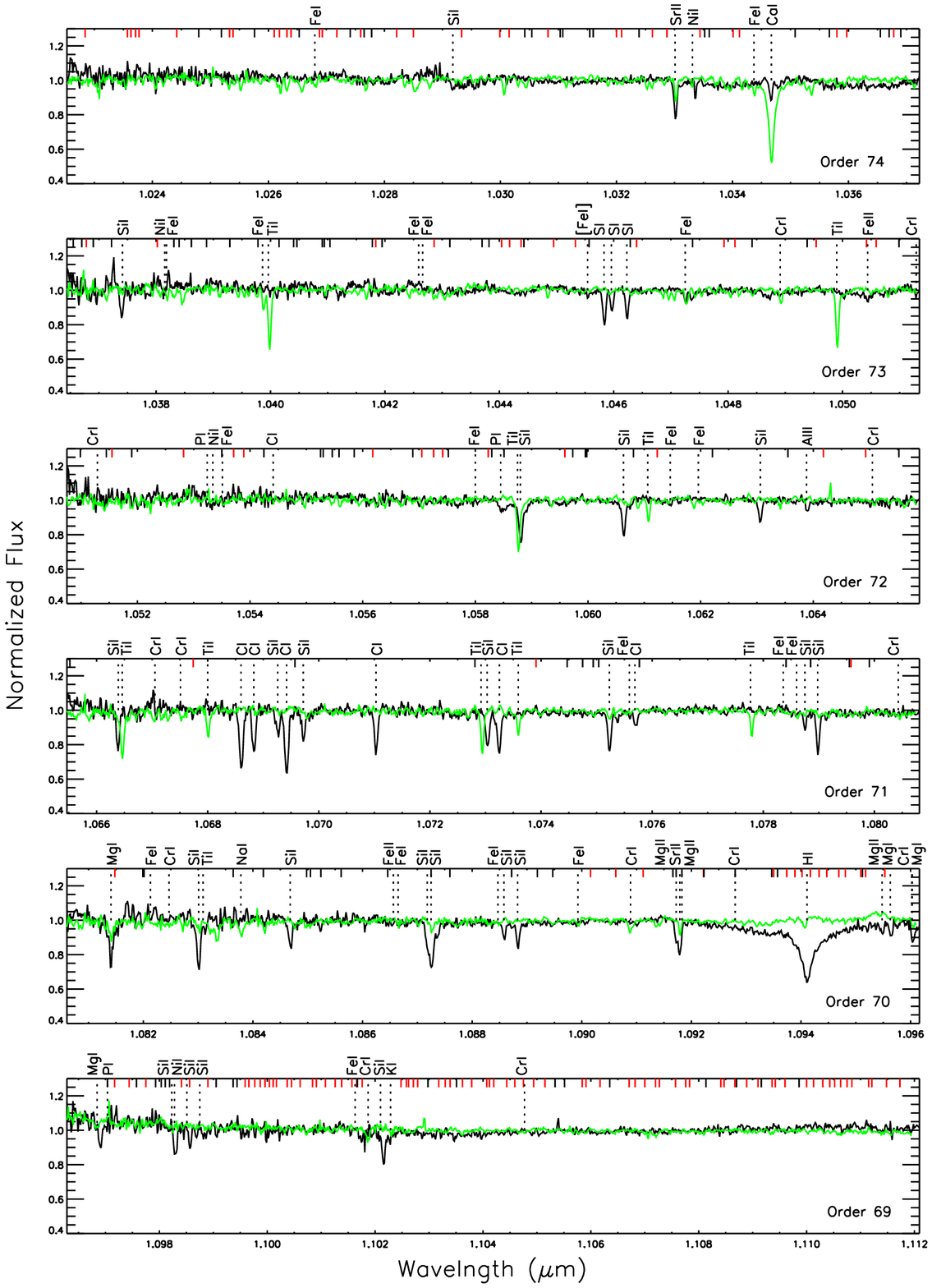}\end{center}
\figurenum{5b}
\caption{(Continued) \label{fig:fmdwarf_pg2}}
\end{figure}

\end{document}